\pdfoutput=1

\documentclass[pdflatex,sn-aps]{sn-jnl}

\usepackage{graphicx}
\usepackage{multirow}
\usepackage{amsmath,amssymb,amsfonts,amsbsy}
\usepackage{mathrsfs}
\usepackage{appendix}
\usepackage{textcomp}
\usepackage{manyfoot}
\usepackage{booktabs}
\usepackage{etoolbox}
\usepackage{csquotes}
\usepackage{xcolor}
\usepackage{cleveref}
\usepackage{csvsimple}
\usepackage{longtable}
\usepackage{url}
\usepackage{hypcap}
\usepackage{hyperref}
\usepackage{natbib}
\usepackage{wrapfig}
\usepackage{amsthm}
\usepackage{empheq}
\usepackage{bm}
\usepackage{ulem}
\usepackage{cancel}

\usepackage{epstopdf}
\usepackage{tabularx}
\usepackage{pdfpages}
\usepackage[T1]{fontenc}
\usepackage[utf8]{inputenc}
\usepackage[activate={true,nocompatibility},final,tracking=true,kerning=true,spacing=true]{microtype}
\jyear{2022}%

\theoremstyle{thmstyleone}%
\theoremstyle{thmstyletwo}%

\theoremstyle{thmstylethree}%

\begin{document}

\title{NPSA: Nonparametric Simulated Annealing for Global Optimization}


\author*[1,2]{\fnm{Rong} \sur{Chen}}\email{chen.rong@asu.edu}

\author[1,3]{\fnm{Alan} \sur{ Schumitzky}}\email{schumitzky@gmail.com}

\author[4]{\fnm{Alona} \sur{Kryshchenko}}\email{alona.kryshchenko@csuci.edu}

\author[1]{\fnm{Julian D.} \sur{Otalvaro}}\email{jotalvaro@chla.usc.edu}

\author[1]{\fnm{Walter M.} \sur{Yamada}}\email{wyamada@chla.usc.edu}

\author[1,5]{\fnm{Michael N.} \sur{Neely}}\email{mneely@chla.usc.edu}

\affil[1]{\orgdiv{Laboratory of Applied Pharmacokinetics and Bioinformatics}, \orgname{Children's Hospital Los Angeles, University of Southern California}, \orgaddress{ \city{Los Angeles}, \postcode{90027}, \state{California}, \country{USA}}}

\affil[2]{\orgdiv{Department of Physics}, \orgname{Arizona State University}, \orgaddress{ \city{Tempe}, \postcode{85287}, \state{Arizona}, \country{USA}}}

\affil[3]{\orgdiv{Department of Mathematics}, \orgname{University of Southern California}, \orgaddress{\city{Los Angeles}, \postcode{90089}, \state{California}, \country{USA}}}

\affil[4]{\orgdiv{Department of Mathematics}, \orgname{California State University Channel Islands}, \orgaddress{\city{Camarillo}, \postcode{93012}, \state{California}, \country{USA}}}

\affil[5]{\orgdiv{Pediatric Infectious Diseases}, \orgname{Children's Hospital Los Angeles, University of Southern California}, \orgaddress{ \city{Los Angeles}, \postcode{90027}, \state{California}, \country{USA}}}


\abstract{
In this paper we describe NPSA,
the first parallel nonparametric global maximum likelihood optimization algorithm using simulated annealing (SA).
Unlike the nonparametric adaptive grid search method NPAG,
which is not guaranteed to find a global optimum solution,
and may suffer from the curse of dimensionality,
NPSA is a global optimizer and it is free from these grid related issues.
We illustrate NPSA by a number of examples including a pharmacokinetics (PK) model for Voriconazole
and show that
NPSA may be taken as an upgrade to the current grid search based nonparametric methods.
}




\maketitle


\section{Introduction}

Among maximum likelihood methods for population modeling of pharmacokinetic (PK) and pharmacodynamic (PD) systems,
parametric and nonparametric algorithms are two of the most common approaches.

Parametric methods assume the shape (usually gaussian) of the distribution of the parameter values,
and try to find those values by maximizing the likelihood.
The popular parametric approaches are usually based on
direct optimization (ICON's NONMEM program \cite{NONMEM,NONMEMwiki}),
expectation maximization (EM) method \cite{Dempster1977a,schumitzky1995algorithms,walker1996algorithm} and Monte Carlo techniques,
such as
ADAPT from the University of Southern California \cite{ADAPT},
S-ADAPT \cite{sadapt},
Quasi-random parametric EM (QRPEM) \cite{QRPEM} in Certara's Phoenix software,
and Stochastic approximation EM (SAEM) \cite{SAEM1999,kuhn2004coupling,KUHN20051020,lavielle2014improved} in Lixoft's Monolix software.

Nonparametric methods do not assume the shape of the distributions of the parameter values.
Instead for maximum likelihood, they consider all distributions on the parameter space and determine which distribution has the highest likelihood.
The most commonly used free nonparametric algorithms today is the nonparametric adaptive grid  NPAG \cite{yamada2020npag} in the Pmetrics R package \cite{neely2012Pmetrics} package.

However, the NPAG method is sensitive to initial conditions,
because it is not a global optimization algorithm.
NPAG is guaranteed to only converge to a local maximum of likelihood.

A standard method to increase the probability of finding the global maximum is to repeatedly initialize with different conditions.
However,
such methods are computationally formidable and not guaranteed to be successful.

Therefore,
developing efficient global maximum likelihood methods
which are not sensitive to initial conditions is an important task.
By using a global optimization algorithm,
one only needs to run it once and obtain the global optimal solution.

In this paper,
we focus on this task for nonparametric methods.
We developed the first parallel nonparametric global maximum likelihood algorithm to use simulated annealing (SA) \cite{Kirkpatrick1983,Corana1987,Goffe1994}.

SA is an elegant stochastic metaheuristic global optimization method for large scale nonlinear optimization problems.
It can deal with both continuous and discrete variable optimization problems.
SA is particularly useful when the objective function contains a large number of local optima
and/or difficult derivatives.
Its implementation is straightforward,
and since the seminal papers by Kirkpatrick et al. \cite{Kirkpatrick1983,Kirkpatrick1984SA},
SA quickly became one of most widely used global optimization algorithms.
There are numerous books discussing SA \cite{VanLaarhoven1987SA,Aarts1989SAbook,Chibante2010SAbook,Ingber2012SAbook},
and its application as a global optimization technique \cite{Press1996FortranNR,Zhigljavsky2007SGObook,Pham2012SAbook,Kochenderfer2019book,Dzemyda2006SGObook}.
SA has been proven to be an extremely useful method across disciplines.
For example,
one can find various applications of SA
in nuclear physics \cite{AuPRC2005a},
in the physical design of computers \cite{Kirkpatrick1983,Rutenbar1989SA},
in combinatorial optimization problems \cite{Kirkpatrick1983},
in neural networks \cite{Goffe1994,Cohen1994MSthesis},
in optimal route detection \cite{Grabusts2019},
in complex portfolio selection problems \cite{Crama2003SA,Busetti2003SA},
and in medical field, such as
in external beam radiation therapy and high dose-rate brachytherapy, see Henderson et al. in Chapter 3 in \cite{Chibante2010SAbook}.

We named our new nonparametric simulated annealing algorithm NPSA.
The paper is organized as follows.
In Sec. \ref{secMethods},
we first briefly introduce maximum likelihood estimate (MLE) for the nonparametric case.
Next, we introduce the idea of simulated annealing,
followed by describing the idea of NPSA, how it works, and the implementation of NPSA code.
We discuss the costs of NPAG and NPSA algorithms in Sec. \ref{secCost},
and point out how to use NPSA-OSAT, namely the one-subject-a-time (OSAT) method of NPSA,
to significantly speedup NPSA.
In Sec. \ref{secResults},
we illustrate NPSA with a number of examples including a pharmacokinetic model for Voriconazole \cite{Neely2015a} and compare NPSA with NPAG.
We show that NPSA can consistently achieve greater likelihood than NPAG
for similar run times.
In Sec. \ref{secSummary}, we conclude the paper and consider possible future directions.

\section{Methods}
\label{secMethods}

\subsection{Maximum Likelihood Estimate}
\label{secMLE}

The population analysis problem can be stated as follows:
let $\bm{Y}_1, \bm{Y}_2 ..., \bm{Y}_n$ be a sequence of independent but not necessarily identically distributed random vectors constructed from one or more observations from each of $n$ subjects in the population.
The $\{ \bm{Y}_i \}$ are observed.
Let $\bm{\theta}_1, \bm{\theta}_2, ..., \bm{\theta}_n$ be a sequence of independent and identically distributed random vectors belonging to a compact subset $\bm{\Theta}$ of Euclidean space with common but unknown distribution $f$.
The $\{\bm{\theta}_i\}$ are not observed.
It is assumed that the conditional densities $p(\bm{Y}_i \vert \bm{\beta}, \bm{\theta}_i)$ are known, for $i = 1,...,n$, where $\bm{\beta}$ is an unknown vector in a set $\textbf{B}$.
The probability of $\bm{Y}_i$ given $\bm{\beta}$ and $f$ is
$
p(\bm{Y}_i \vert \bm{\beta}, f)
=
\int p(\bm{Y}_i \vert \bm{\beta}, \bm{\theta}_i)
f(\bm{\theta}_i) d \bm{\theta}_i.
$
Because of independence of the $\{ \bm{Y}_i \}$,
the probability of the $\{ \bm{Y}_i \}$ given $\bm{\beta}$ and $f$ is given by the likelihood,
\begin{eqnarray}
L(\phi) = \prod_{i=1}^n
\int p(\bm{Y}_i \vert \bm{\beta}, \bm{\theta}_i)
f(\bm{\theta}_i)
d \bm{\theta}_i,
\label{L}
\end{eqnarray}
where $\phi$ is given by
\begin{align}
\label{phi_general}
 \phi =   \{ \bm{\beta}, f \},
\end{align}

In the context of so-called mixed-effects problems, the vector $\bm{\beta}$ would describe the
``Fixed'' effects and the $\{\bm{\theta}_i\}$ would describe the ``Random'' effects.

The population analysis problem is to maximize the likelihood function $L(\phi)$ with respect to all parameters $\bm{\beta}$ in $\textbf{B}$ and all density functions $f$ on $\bm{\Theta}$.

The maximum likelihood problem as stated above is infinite dimensional.
The important theorems of Mallet \cite{Mallet1986a} and Lindsay \cite{Lindsay1983a} reduce this problem to finite dimensions.
It was proved under simple hypotheses on the conditional densities $p(\bm{Y}_i \vert \bm{\beta}, \bm{\theta}_i)$,
that for fixed $\bm{\beta}$,
the optimal density $f$ could be found in the space of discrete densities with no more than $n$ support points, where n is the number of subjects.
This is a remarkable conclusion because now $K$ is a bounded value,
which makes all the numerical computations feasible.
Therefore $f$ can be written as a weighted sum of delta functions,

\begin{equation}
f(\bm{\theta})
= \sum_{k=1}^{K} w_k \delta (\bm{\theta} - \bm{\mu}_k  ),
\label{nonparametric_p}
\end{equation}
where $K \leq n$, $w_k \geq 0$ and $\sum_{k=1}^{K} w_k = 1$.
In Eq. (\ref{nonparametric_p}) the weights $w_k$ and the vector $\bm{\mu}_k$ are implicit functions of $\bm{\beta}$.
Usually we call $\bm{\mu}_k$ a support point \cite{yamada2020npag}.

Therefore Eq.(\ref{L}) becomes
\begin{eqnarray}
L(\phi) =
\prod_{i=1}^n
\sum\limits_{k=1}^{K}
w_k p(\bm{Y}_i \vert \bm{\beta}, \bm{\mu}_k),
\label{nonparametric_L}
\end{eqnarray}
where
\begin{align}
\phi = \{ \bm{\beta}, f \} =  \{ \bm{\beta}, (w_k, \bm{\mu}_k), k=1,...,K  \}.
\label{nonparametric_phi}
\end{align}

The quantity $p(\bm{Y}_i\vert \bm{\beta}, \bm{\mu}_k)$
can be taken as the $(i,k)$ element of the $n$ by $K$ likelihood matrix,
\begin{equation}
\left (
\begin{array}{ccccc}
p(\bm{Y}_1\vert\bm{\beta}, \bm{\mu}_1) & p(\bm{Y}_1\vert\bm{\beta}, \bm{\mu}_2) & ... & p(\bm{Y}_1\vert\bm{\beta}, \bm{\mu}_{K-1}) & p(\bm{Y}_1\vert\bm{\beta}, \bm{\mu}_K)  \\
p(\bm{Y}_2\vert\bm{\beta}, \bm{\mu}_1) & p(\bm{Y}_2\vert\bm{\beta}, \bm{\mu}_2) & ... & p(\bm{Y}_2\vert\bm{\beta}, \bm{\mu}_{K-1}) & p(\bm{Y}_2\vert\bm{\beta}, \bm{\mu}_K)  \\
\vdots & \vdots & \ddots & \vdots & \vdots  \\
p(\bm{Y}_{n-1}\vert\bm{\beta}, \bm{\mu}_1) & p(\bm{Y}_{n-1}\vert\bm{\beta}, \bm{\mu}_2) & ... & p(\bm{Y}_{n-1}\vert\bm{\beta}, \bm{\mu}_{K-1}) & p(\bm{Y}_{n-1}\vert\bm{\beta}, \bm{\mu}_K)  \\
p(\bm{Y}_n\vert\bm{\beta}, \bm{\mu}_1) & p(\bm{Y}_n\vert\bm{\beta}, \bm{\mu}_2) & ... & p(\bm{Y}_n\vert\bm{\beta}, \bm{\mu}_{K-1}) & p(\bm{Y}_n\vert\bm{\beta}, \bm{\mu}_K)  \\
\end{array}
\right ) \,.
\label{psi_mat}
\end{equation}

Note that usually it is the log likelihood function $L(\phi)$ that is calculated,
which can be written as
\begin{align}
\ln L(\phi) = \sum\limits_{i=1}^{n}\ln \left( N_i \right),
\label{LL}
\end{align}
where we define $N_i$ as,
\begin{align}
N_i \equiv \sum\limits_{k=1}^{K} w_k n_{ik},
\label{Ni}
\end{align}
and
$n_{ik}$ is simply the $(i,k)$ element of the $n$ by $K$ likelihood matrix Eq.(\ref{psi_mat}),
\begin{align}
n_{ik} \equiv
p(\bm{Y}_i\vert \bm{\beta}, \bm{\mu}_k).
\label{nonparametric_n_ik}
\end{align}

In many cases, the $m_i$-dimensional observation vector for the $i$th individual $\bm{Y}_i=(Y_{1 i}, ..., Y_{m_i i})$ is sampled from a Gaussian distribution such that,
\begin{equation}
\bm{Y}_i \vert \bm{\beta}, \bm{\theta}_i
\sim
N(\bm{h}_i(\bm{\theta}_i),\bm{G}_i(\bm{\beta},\bm{\theta}_i)),
~~ i=1,...,n
\label{stage1}
\end{equation}
where
$\bm{h}_i(\bm{\theta}_i)$ is the function defining the PK/PD model, and
$\bm{G}_i(\bm{\beta}, \bm{\theta}_i)$ is a positive definite covariance matrix ($\bm{G}_i \in R^{m_i \times m_i}$).
For the examples shown in this paper, we use the following case which is important \cite{wang2007nonlinear} and commonly used,
\begin{eqnarray}
\bm{G}_i(\bm{\beta}, \bm{\theta}_i) = \sigma^2 \bm{H}_i(\bm{\theta}_i),
\label{stage1_sigma2}
\end{eqnarray}
where $\bm{H}_i(\bm{\theta}_i)$ is a known function and $\bm{\beta} = \sigma^2$.

\subsection{D function}
\label{secDfunc}

For the nonparametric approach,
if $\bm{\beta}$ is known, and $\phi = f$ is any density,
we can use Fedorov D function $D(\phi)$ \cite{Fedorov_D, Lindsay1983a},
\begin{align}
D(\phi) = n \ln \left[1 + \frac{\max\limits_{\bm{\theta}} D(\bm{\theta},\phi)}{n} \right]
\label{Dfunc}
\end{align}
where
\begin{align}
D(\bm{\theta},\phi)
&= \sum_{i=1}^{n} \frac{ p(\bm{Y}_i\vert \bm{\beta}, \bm{\theta}) }
{ \int p(\bm{Y}_i\vert \bm{\beta}, \bm{\theta}) f(\bm{\theta}) d \bm{\theta} } - n \\
&= \sum_{i=1}^{n} \frac{p(\bm{Y}_i\vert \bm{\beta}, \bm{\theta})}{N_i} -n ,
\label{Dfunc1}
\end{align}
to estimate the maximum distance between the true global maximum log likelihood $L(\phi_{ML})$ and the current obtained  log likelihood $L(\phi)$,
\begin{align}
L(\phi_{ML}) - L(\phi) \leq D(\phi).
\label{LLmaxdiff}
\end{align}

We emphasize that this D function is for fixed known $\bm{\beta}$.
When there is no $\bm{\beta}$ in the problem, the symbol $\bm{\beta}$ can be deleted in the D function.
For fixed $\bm{\beta}$ there is an elegant theory which revolves around the directional derivative of $L(\phi)$ with respect to the distribution $f$.
This is the theory of the D function of Fedorov \cite{Fedorov_D}.
In this paper, for all the examples
except the example in Sec. \ref{secEx1}, there is no $\bm{\beta}$.

If $\phi = \phi_{ML}$, $D(\phi)$ in Eq.(\ref{LLmaxdiff}) should be zero.
The closer to zero $D(\phi)$ is, the closer the current solution $\phi$ is to the true global optimum solution $\phi_{ML}$.
However, as is shown in Eq.(\ref{Dfunc}), calculating $D(\phi)$ requires maximizing $D(\bm{\theta},\phi)$.
If the algorithms used to maximize $D(\bm{\theta},\phi)$ failed to find the global maximum of $D(\bm{\theta},\phi)$, the evaluation of $D(\phi)$ will be smaller than what it should really be.
Due to this limitation, a small $D(\bm{\theta},\phi)$ does not necessary means the global optimum is reached.
In this paper, simulated annealing is used to calculate $D(\phi)$ after a candidate distribution has been determined.
In principle, different global optimum algorithms can be used to double check $D(\phi)$.
Nonetheless,
a large $D(\bm{\theta},\phi)$ for sure indicates the current solution $\phi$ is not the global optimum solution.

\subsection{Simulated Annealing}
\label{secSA}

NPSA is an application of SA under a nonparametric framework.
It uses SA to find the global MLE of $\phi$.
We will show how to use simulated annealing (SA) \cite{Kirkpatrick1983,Corana1987,Goffe1994}
to find the global MLE of $\phi$.
Before go into NPSA, we begin by briefly introducing SA.

\subsubsection{Mechanism}
 SA uses a computer to simulate a physics process
which naturally evolves an objective function towards the global optimum solution.
The name of SA comes from annealing in metallurgy,
and the idea of SA originated from statistical mechanics in physics.
The mechanism of SA can be described as follows.

According to statistical physics,
when a system at temperature $T$ is in equilibrium,
the probability of the system at state $s$ is given by (in natural units),
\begin{align}
p(s) = \frac{e^{-\frac{E(s)}{T}}}{Z}
\label{sa_prob}
\end{align}
where $E(s)$ is the energy (the objective function) of the system in state $s$,
and $Z$ is called partition function and it is a normalization factor, i.e.,
$Z = {\sum\limits_s e^{-\frac{E(s)}{T}}   }$.
At each temperature, when in equilibrium, the states of the system will be distributed according to Eq.(\ref{sa_prob}).

From Eq.(\ref{sa_prob}) we can see that, when the temperature $T$ is very high, all the states with corresponding energy $E(s)$ have almost the same probability. In other words, the system have almost equal chance to be at any of the states.
When the temperature $T$ approaches zero,
only those states with the lowest energy will have significantly higher probability than other states.
In other words, the system will be occupied only by the states with the lowest energy,
which is also called ground state energy.

If we let the system begin from a high temperature,
and slowly decrease the temperature $T$,
let the system form equilibrium at each temperature,
finally, as the temperature $T$ approaches zero,
the system will be occupied by ground state with ground state energy.
This process is similar with the annealing process in metallurgy,
and the task of SA is to use computer to simulate such a process and find the ground state and the corresponding ground state energy.

In fact,
the energy $E(s)$ is actually the objective function to be minimized in optimization problems.
The ground state energy means the global optimal value of the object function,
and the ground state means the global optimal parameters in the object function.
In the current NPSA algorithm, such an objective function is the negative log likelihood.

\subsubsection{Algorithm}

To simulate the SA process,
we begin from a reasonably high temperature and a state $s$,
and we use Metropolis algorithm \cite{Ceperley95a} to form the target distribution which is $p(s)$ in
Eq.(\ref{sa_prob}).
To do so, we propose a new state $s'$ according to proposal probability $T(s\rightarrow s')$, and judge if it is accepted by the acceptance rate $A(s \rightarrow s')$, which is typically chosen as
\begin{equation}
A(s\rightarrow s')=
\min \left[  1, \frac{p(s')T(s'\rightarrow s)}{p(s) T(s\rightarrow s')} \right]
\label{Accptequation}
\end{equation}
in order to satisfy the detailed balance condition \cite{Kalos1986a}, i.e.,
$
p(s')T(s'\rightarrow s)A(s'\rightarrow s)=p(s) T(s\rightarrow s')A(s\rightarrow s')
$.
Since we usually choose a proposal probability such that $T(s\rightarrow s')=T(s' \rightarrow s)$,
The acceptance rate Eq.(\ref{Accptequation}) becomes,
\begin{align}
A(s \rightarrow s') =
\min \left[1,
\frac{p(s')}{p(s)}
 \right].
\label{sa_ar}
\end{align}
The Metropolis algorithm in SA is described as follows.
\begin{description}
  \item[\textbf{Step 1}] For the current state $s$, propose a new state $s'$.
  \item[\textbf{Step 2}] For the proposed new state $s'$, use Eq.(\ref{sa_ar}) to calculate the acceptance rate $A(s\rightarrow s')$.
  \item[\textbf{Step 3}] Generate a uniform random number $x$ which ranges from 0 to 1.
  If $x \leq A(s\rightarrow s')$, the new state $s'$ is accepted, so set $s'$ as the current state $s$, namely $s = s'$, and keep $s'$ as a sample, then go to step 1 and continue.
  Else, the new state $s'$ is rejected which means the state $s$ is accepted and kept as the sample,
  so set $s$ as our current state $s$, then go to step 1 and continue.
\end{description}
After iterating the Metropolis algorithm enough times,
all those `current state' $s$ in step 1 will be distributed according to the target distribution $p(s)$.

After we form the distribution of $p(s)$,
we begin from the state with the lowest energy at the current temperature $T$,
and we lower the temperature, and repeat the Metropolis process at this lowered temperature.
The whole simulated annealing process terminates when the lowest energies at several successive temperatures
become almost the same and according to certain stopping criterion \cite{Corana1987}.
We take the lowest energy as the ground state energy (the value of the global optimal objective function), and its corresponding state as the ground state (the parameters in the global optimal objective function).

Note that,
we can see from the Metropolis algorithm Eq.(\ref{sa_ar}),
SA does not require calculating gradient
which may encounter numerical instability issues.
Furthermore, SA is not a greedy method,
in the sense that it can accept some proposed states $s'$ whose probability $p(s')$ is smaller than the old state's $p(s)$.
This feature together with the temperature deceasing scenario,
allows SA to escape from local optima and explore the parameter space systematically,
making SA a robust global optimization algorithm.
Next we will show how to use SA for nonparametric maximum likelihood approach, i.e., NPSA.

\subsection{NPSA Algorithm}
\label{secNPSA}

\subsubsection{Formulism}

NPSA is the nonparametric algorithm which uses SA to find the global MLE of $\phi$ (state) which minimize the objective function (energy).
For the convenience of computing,
usually it is the log likelihood (LL) function $L(\phi)$ in Eq.(\ref{L}) that is calculated and maximized.
In this paper, we take averaged  log likelihood $-\frac{\ln L(\phi)}{n}$ as the objective function to be minimized
\footnote{
In general the objective function does not have to be $-\frac{\ln L(\phi)}{n}$ alone,
some penalty functions can be used as well in order to favor certain $\phi$, such as those used in solving the classic travelling salesman problem, etc \cite{Kirkpatrick1983}.
We may also add certain diversity function in the objective function in order to, for example, better catch outliers, etc.
},
which is more bounded than  log likelihood $-{\ln L(\phi)}$.
So, the correspondence between the physics system and the optimization problem is,
\begin{align}
s &= \phi, \\
E(s)  &= -\frac{\ln L(\phi)}{n},  \\
A(s \rightarrow s') &=
\min \left[1,
\frac{e^{\frac{\ln L(\phi')}{nT}}}{e^{\frac{\ln L(\phi)}{nT}}}
 \right],
\label{sa_ar_npsa}
\end{align}
where the log likelihood function $L(\phi)$ is calculated as shown in Eq.(\ref{LL}).

\subsubsection{Algorithm}
\label{SecAlgo}

NPSA is an application of SA on nonparametric problem.
To ensure the robustness and make NPSA generalizable,
its main SA part is based on \cite{Goffe1994},
in the sense that,
we propose the new $\phi'$ by moving each element in $\phi$ one by one,
and the jump between $\phi$ and the new proposed $\phi'$ is controlled by the step size which ensures the acceptance rate Eq.(\ref{sa_ar_npsa}) between $40\%$ and $60\%$.
The main difference of our NPSA from the code in \cite{Goffe1994} is that, we customized and parallelized the SA algorithm for nonparametric problems, for example, the calculation of the objection function is parallelized.
Next, we describe how NPSA works.

In NPSA, all the $\bm{\beta}$ and $\bm{\mu}_k$ in $\phi$ are bounded by user-provided lower and upper bounds.
Each of the $w_k$ needs to be bounded between 0 and 1.
There can be three choices of $\phi$ in implementing NPSA, and we list them and their advantages and disadvantages as follows.
\begin{description}
  \item[\textbf{Choice 1}. ]
  Setting the number of support points $K$ the same as the number of subjects $n$, and take $\phi$ as,
  \begin{align}
\phi = \{ \bm{\beta}, \bm{\mu}_k, k=1,...,n  \},
\label{nonparametric_phi1}
\end{align}
which means SA only vary $\sigma$ and all the $\bm{\mu}_k$,
while all the optimal $w_k$ are calculated once $\{ \bm{\beta}, \bm{\mu}_k, k=1,...,n  \}$ are given.
For example, by using Eq.(\ref{nonparametric_p}),
each of the optimal $w_k$ can be obtained from the solution of the iteration relation \cite{RPEM2022},
\begin{align}
w_k^{(r+1)}
=\frac{1}{n}  \displaystyle  \sum\limits_{i=1}^{n}
\frac{w^{(r)}_k p(\bm{Y}_i\vert {\bm{\beta},\bm{\mu}}_k)}
{ \displaystyle \sum\limits_{k=1}^{K} w^{(r)}_k p(\bm{Y}_i\vert \bm{\beta},\bm{\mu}_k) },
\label{wk_delta}
\end{align}
where superscript $(r)$ denotes the $r$th iteration.
The optimum weights $w_k$ can also be obtained from the Primal-Dual Interior-Point method (PDIP) in \cite{yamada2020npag}, and from the method described in \cite{Wang2007fastNPML}.

The advantage of this choice is that,
it can be very accurate since the optimal $w_k$ is calculated.
However the disadvantage is that,
since every time any one of the components in $\phi$ is varied,
each of the optimal $w_k$ need to be calculated again which requires an additional  $\mathcal{O}(n^2)$ operations,
this will slow down the computation noticeably.
Due to this speed disadvantage, choice 1 is not implemented in NPSA.

\item[\textbf{Choice 2}. ]
Take $\phi$ as,
\begin{align}
\phi &= \{ \bm{\beta}, w_k = \frac{1}{K}, \bm{\mu}_k, k=1,...,K  \},
\label{nonparametric_phi2}
\end{align}
which means all the weights $w_k$ are treat as the same as $1/K$.
In this case, $K$ can be bigger than the number of subjects $n$.
In fact, when $K$ is big enough, the support points in Eq.(\ref{nonparametric_phi2}) can be approximately combined as at most $n$ support points with corresponding weights.
Once we use SA find the optimal $\phi$,
we can begin from $w_k^{(0)}=1/K$ and use the iteration relation Eq.(\ref{wk_delta})
to get the optimal weights for each support points.

The advantage of this choice that, there is no need to calculate the optimal weights $w_k$ during SA
since we can simply use more support points with the same weights instead.
Because of this, the code can be relatively easy to parallelize.
The disadvantage is that we may need more than $n$ support points,
this will slow down the computation.

\item[\textbf{Choice 3}. ]
This is the default choice.
It sets the number of support points $K$ the same as the number of subjects $n$,
such that
\begin{align}
\phi = \{ \bm{\beta}, (w_k, \bm{\mu}_k), k=1,...,n  \},
\label{nonparametric_phi_choice3}
\end{align}
which means besides varying the support points $\bm{\mu}_k$ and $\sigma$,
SA will vary all the weights $w_k$ as well,
and find the global optimal $\phi$ which minimize $-\ln L(\phi)$.

Note that, considering the normalization of the weights $w_k$ which is Eq.(\ref{normalization_wk}), for choice 3 only $n-1$ weights are independent.
So, in NPSA,
we propose the new weights in $\phi'$ by varying the first $n-1$ weights from $w_1$ to $w_{n-1}$ one by one (they are bounded between 0 and 1),
and the last weight $w_n$ is calculated by $w_n=1- \sum\limits_{k=1}^{n-1} w_k$.
If $w_K<0$, we reject that $\phi'$.

In principle, choice 3 is the ``one-size-fits-all'' solution,
because it stochastically varies both the weights and the support points in order to find the global maximum likelihood solution.
Also, we find choice 2 and choice 3 usually work equally well and the computation time is similar. Therefore, we take choice 3 as the default option for NPSA.
\end{description}

Regardless of choice 1, 2, or 3,
when we propose $\phi'$ from $\phi$,
the most time consuming part in NPSA is when any support point $\bm{\mu}_k$
is changed and becomes $\bm{\mu}'_k$.
The new $\bm{\mu}'_k$ requires re-calculating the $k$th column in the matrix Eq.(\ref{psi_mat}) which involves time-consuming ODEs to solve.
So this part will benefit most from parallelization.
Furthermore, although usually SA is a serial algorithm,
in the nonparametric case we can indeed propose one element in each $\bm{\mu}_k$ at a time independently,
and then judge if each element is accepted or not.
This allows us to calculate the new proposed matrix of
$p(\bm{Y}_i\vert\bm{\beta}, \bm{\mu}_k)$ in advance in parallel.
Therefore, we parallelize this part in NPSA using MPI and describe it in Sec. \ref{SecImplementation}.

\subsubsection{Implementation}
\label{SecImplementation}

Since choice 3 in Sec. \ref{SecAlgo} is the most comprehensive and the default choice of NPSA, we take it as an example
and describe the implementation of NPSA algorithm (the implementation of choice 1 and 2 are similar).
We begin by introducing some notations.

\begin{description}
\item[\textbf{Notation 1}.]
For choice 3,
$\phi = \{ \bm{\beta}, (w_k, \bm{\mu}_k), k=1,...,n  \}$, so the matrix of $p(\bm{Y}_i\vert\bm{\beta}, \bm{\mu}_k)$ is n by n.
The dimension of $\sigma$ is 1 if we look at Eq.(\ref{stage1_sigma2}),
and we denote the dimension of each support point $\bm{\mu}_k$ as $d_{\mu}$,
the number of $\bm{\mu}_k$ is $n$,
the number of independent weights $w_k$ is $n-1$.
We denote the total dimension of $\phi$ as $d_{tot}$,
which therefore becomes
$d_{tot} = 1+(n-1)+n\times d_{\mu}$.
We define a one-dimension integer array which contains $d_{tot}$-element,
and we call it $q-$array
which contain integer from 1 to $d_{tot}$.
The order of the elements in $\phi$ is fixed, however we can use $q-$array to shuffle the elements in $\phi$. For example, if the $i$th element in $q-$array is 7, it means the 7th element in $\phi$.
Both $q-$array and $\phi$ have $d_{tot}$ elements, so there is a one-to-one correspondence.

\item[\textbf{Notation 2}.]
We define the ``effective total dimension'' as $d_{eff} = d_{\mu} + 1 + (n-1)$,
and we define a one-dimensional integer array containing $d_{eff}$ elements, which are integers from 1 to $d_{eff}$, and we call this array $d_{eff}-$array.
For example, if the $i$th element in $d_{eff}-$array is 5,
it is associated with the $5$th dimension in $\bm{\mu}_k$.
The last $1+(n-1)$ elements associate with $\sigma$ and the $n-1$ independent $w_k$.
This array will also be shuffled which allows us to propose the elements in $\phi$ more randomly.

\item[\textbf{Notation 3}.]
We denote $N_{proc}$ as the number of CPU cores.
We denote the first CPU core as the master core.
Each of the CPU cores is associated with almost the same number of elements
\footnote{
Each of the CPU cores is associate with $n^2/N_{proc}$ elements if $\textrm{mod} (n^2,N_{proc})=0$.
Otherwise either $\textrm{int} (n^2/N_{proc})$ or $\textrm{int} (n^2/N_{proc})+1$ elements.
So the number of elements associate with each CPU cores are at most differ by 1.
\label{footnote2}}
in the n by n matrix of $p(\bm{Y}_i\vert\bm{\beta}, \bm{\mu}_k)$.
This makes the workload of each CPU cores almost the same,
when solving ODEs in calculating the matrix of $p(\bm{Y}_i\vert\bm{\beta}, \bm{\mu}_k)$.

\item[\textbf{Notation 4}.]
We denote $N_t$, $N_s$ which have the same meaning as in \cite{Goffe1994}.
Usually $N_s$ is set to 20, $N_t$ is set to 10.
Once each the $d_{tot}$ elements in $\phi$ has been proposed one by one once,
we call it a ``sweep''.
After $N_s$ sweep, we adjust the proposal step size
\footnote{
NPSA is in fact adaptive in the sense that its step size is adjusted during the process, in a way similar with the adaptive grid part in NPAG.
}
for each of the $d_{tot}$ elements in $\phi$ from the acceptance rate in the Metropolis process according to \cite{Corana1987,Goffe1994}, to keep the acceptance rate between $40\%$ and $60\%$.
After adjusting the step size $N_t$ times (so totally $N_s \times N_t$ sweeps), we decrease the temperature according to certain scenario.
\end{description}

The implementation of NPSA is described as follows.
\begin{description}
\item[\textbf{Step 0}. ]
The master core calculates each of the elements in the $n$ by $n$ matrix of $p(\bm{Y}_i\vert\bm{\beta}, \bm{\mu}_k)$, given the initial $\phi$.
The initial temperature $T$ can be set as the absolute value of the corresponding object function $\vert E(\phi) \vert$.
Empirically, we found for most models and data with 50 subjects,
the log likelihood is about the order of $10^3$.
So averaged log likelihood per subject is around an order of $10^2$.
In this paper, we set initial temperature about 60.

\item[\textbf{Step 1}. ]
  At the current temperature $T$, the master core shuffles the $q-$array and the $d_{eff}-$array.
  Then the master core broadcasts the shuffled $q-$array and the $d_{eff}-$array to all the CPU cores.
  The shuffle is to ensure that we are proposing each elements in $\phi$ randomly,
  so it would not introduce implicit proposal probability which may cause the Metropolis judgement Eq.(\ref{sa_ar_npsa}) to be invalid.

\item[\textbf{Step 2}. ]
  From index 1 to $d_{eff}$ in $d_{eff}-$array,
  all the CPU cores loop over the elements in $d_{eff}-$array.
  For example, in the loop, for the $i$th index, if the $i$th element in $d_{eff}-$array is smaller or equal to $d_{\mu}$, then go to step 3.
  Otherwise, we only use the master core,
  because in such cases no ODEs need to be solved,
  the master core is fast enough.
  We propose the corresponding $\sigma'$ or $w'_k$ and judge if they are accepted or not using Eq.(\ref{sa_ar_npsa}) and save the corresponding $p(\bm{Y}_i\vert\bm{\beta}, \bm{\mu}_k)$ matrix and then go to step 5.

  \item[\textbf{Step 3}. ]

  In this case, $i$th element in $d_{eff}-$array is smaller or equal to $d_{\mu}$, which means
  the master CPU core needs to do a loop,
  which propose to move one of the elements in $\bm{\mu}_k$ for each support point,
  according to the shuffled $q-$array.

  About this loop, its index $j$ is loop from 1 to $d_\mu$.
  For example, if $n=2$, and the parameters are $K$ and $V$ so $d_\mu=2$,
  there is $\sigma$ and an independent weight $w_1$,
  then $d_{tot}=1+(2-1)+2 \times 2 = 6$.
  The original $q-$array before shuffle is $(1,2,3,4,5,6)$.
  The 1 and 2 in the $q-$array means the $(K,V)$ for the first support point,
  the 3 and 4 in the $q-$array means the $(K,V)$ for the second support point,
  the 5 in the $q-$array means $\sigma$,
  the 6 in the $q-$array means $w_1$.
  Since $q-$array is shuffled, it can be $(5,3,1,6,4,2)$ for example.
  In this example, we delete 5 and 6 in the shuffled $q-$array because they represent $\sigma$ and $w_1$ which is irrelevant here. So we left with $(4,1,2,3)$ which can be interpret as,
  move 4 (the $V$ in support point 2),
  move 1 (the $K$ in support point 1),
  move 2 (the $V$ in support point 1),
  move 3 (the $K$ in support point 2).
  So for support point 1, we move the $K$ first then we move $V$,
  while for support point 2, we move the $V$ first then we move $K$.
  It means that,
  in the loop when the index $j=1$ we propose to move the $K$ in support point 1 and the $V$ in support point 2, when the index $j=2$ we propose to move the $V$ in support point 1 and the $K$ in support point 2.
  Once the loop finished, it means each element in each support point has been proposed.

  During the loop, for a certain index $j$ in the loop,
  each support point has been proposed to be a new support point $\bm{\mu}'_k$,
  and each element in the new $n$ by $n$ matrix of $p(\bm{Y}_i\vert\bm{\beta}, \bm{\mu}'_k)$ needs to be calculated.
  Then the master CPU core distributes the calculation of the $n^2$ elements in $p(\bm{Y}_i\vert\bm{\beta}, \bm{\mu}'_k)$ matrix to all the CPU cores, according to \cref{footnote2} on \Cpageref{footnote2}.
  Once each CPU cores finishes its job, it sends corresponding new elements in $p(\bm{Y}_i\vert\bm{\beta}, \bm{\mu}'_k)$ matrix back to the master CPU core.
  Then the master CPU core goes to step 4.

 \item[\textbf{Step 4}. ]

  With the newly calculated $p(\bm{Y}_i\vert\bm{\beta}, \bm{\mu}'_k)$ matrix and the old $p(\bm{Y}_i\vert\bm{\beta}, \bm{\mu}_k)$ matrix,
  the master CPU core uses the Metropolis algorithm Eq.(\ref{sa_ar_npsa}) to check if the new $\bm{\mu}'_k$ is accepted or not ($\phi$ and $\phi'$ are differ only by $\bm{\mu}_k$ and $\bm{\mu}'_k$), one by one, for all the $n$ support point from $k=1$ to $k=n$.
  If $\bm{\mu}'_k$ is rejected, do nothing.
  If accepted, set $\bm{\mu}'_k$ as $\bm{\mu}_k$, and replace the $k$th column in $p(\bm{Y}_i\vert\bm{\beta}, \bm{\mu}_k)$ matrix
  by the $k$th column in $p(\bm{Y}_i\vert\bm{\beta}, \bm{\mu}'_k)$ matrix.
  Until here, one of the elements in each of the $n$ support points has been proposed.
  Then check if the loop in step 3 is finished, if not go to step 3 and continue the loop; if yes go to step 5.

  \item[\textbf{Step 5}. ]
  Once the master CPU finishes the Metropolis process,
  all the CPU go to step 2 and continue to finish the loop in the $d_{eff}-$array.
  Once the loop in the $d_{eff}-$array is finished, a sweep is finished, all the CPU cores go to step 6.

  \item[\textbf{Step 6}. ]
  Count the number of sweeps.
  If the number of sweeps is less than $N_s$, go to step 1 and continue from there.
  Once the number of sweeps is equal to $N_s$, adjust the proposal step size for each of the $d_{tot}$ elements in $\phi$ according to \cite{Goffe1994}.
  Then set the counter for the number of sweeps to 0.
  With the new proposal step size, all the CPU cores go to step 7.

 \item[\textbf{Step 7}. ]
 Count the number of proposal step size adjustments.
 If this number is less than $N_t$, all the CPU cores go to step 1 and continue from there.
 Once the number of proposal step size adjustments equals $N_t$,
 we pick the optimum $\phi_{opt}$ whose $-\frac{\ln L(\phi_{opt})}{n}$ is the lowest at the current temperature $T$,
 and we check if the stopping criterion \cite{Goffe1994} is satisfied
\footnote{
 We denote $EPS$ as the precision, and $N_{eps}$ as an integer.
At the current temperature $T$,
there is an optimal $\phi_{opt}$ whose $-\frac{\ln L(\phi_{opt})}{n}$ is the lowest.
The stopping criterion of NPSA is that \cite{Goffe1994},
during the Metropolis process, if the final $-\frac{\ln L(\phi)}{n}$ from the last $N_{eps}$ temperatures differ from the corresponding value at the current temperature by less than $EPS$,
and the final $-\frac{\ln L(\phi)}{n}$ at the current temperature
differs from the current optimal $-\frac{\ln L(\phi_{opt})}{n}$ by less than $EPS$, NPSA terminates.
Then the current optimal $\phi_{opt}$ is the global optimal solution.
 },
 if yes the optimum $\phi_{opt}$ is the global optimal solution, report it and the corresponding log likelihood $\ln L(\phi_{opt})$, and NPSA is finished;
 if no then we decrease the temperature (every time we decrease the temperature, we count it as a cycle),
 and the relation between the new temperature $T_{new}$ and the current temperature $T$ can be
\begin{align}
T_{new} = R_t \times T,
\label{SA_T_adjustment}
\end{align}
where $R_t$ is a rate which is usually set between 0.5 and 0.9, and is always less than 1.
Then we begin from the optimum $\phi_{opt}$ with its $p(\bm{Y}_i\vert\bm{\beta}, \bm{\mu}_k)$ matrix,
and we set $T_{new}$ as $T$,
and all the CPU cores go to step 1 and continue from there.

\end{description}

\subsection{The Cost of the NPAG and NPSA algorithms}
\label{secCost}

Both NPSA and NPAG can be efficiently parallelized, therefore they are not only suitable for personal computers but also supercomputer clusters for challenging model and big data sets.

\subsubsection{The cost of NPAG}
\label{secNPAGcost}
In principle, the cost of NPAG is
$\mathcal{O}(n a^d)$ + $\mathcal{O}(n^2 d)$,
where $a$ is a constant,
$d$ is the dimension of $\bm{\theta}$,
$n$ is the number of subjects.
The $\mathcal{O}(n a^d)$ is due to grid search,
as the dimension $d$ grows,
in order to cover the $d-$dimension space evenly, in principle we need to calculate the ODEs for $\mathcal{O}(a^d)$ grid points for each of the $n$ subject, hence $\mathcal{O}(n a^d)$.
In the NPAG algorithm in Pmetrics \cite{neely2012Pmetrics,yamada2020npag},
typically $10^4$ to $10^6$ grid points are used and can usually achieve solid results.
The $\mathcal{O}(n^2 d)$ comes from the fact that,
during the PDIP process usually the number of support points are proportional to $nd$,
and if any of the support point is changed,
it needs to calculate the ODEs for all the $n$ subjects, hence $\mathcal{O}(n^2 d)$ ODEs calls involved.

\subsubsection{The cost of NPSA}
\label{secNPSAcost}
The cost of NPSA is $\mathcal{O}(n^2 d N_t N_s)$.
The $n^2 d$ because it needs to calculate the ODEs for each element in the $n$ by $n$ likelihood matrix Eq.(\ref{psi_mat}), and NPSA moves each of the $d$ dimension in each element one by one, hence $n^2 d$.
The $N_t N_s$ because NPSA will do sweep for $N_t N_s$ times.
There is no grid search term $\mathcal{O}(a^d)$,
so NPSA does not have the dimensionality curse which is a typical issue for grid search methods.

\subsubsection{NPSA-OSAT}
\label{secNPSAOSATcost}
Furthermore, if we use NPSA to find the optimal support point for only one subject,
it only costs $\mathcal{O}(d N_t N_s)$ ODE operations.
So if we find the optimal support point for each subject one by one,
the total cost will be $\mathcal{O}(n d N_t N_s)$ instead of $\mathcal{O}(n^2 d N_t N_s)$.
The NPSA-OSAT is at least $n$ time faster than NPSA,
since it scales NPSA's cost from $\mathcal{O}(n^2)$ down to $\mathcal{O}(n)$.
Once we have all the optimal support points, we can use Eq.(\ref{wk_delta}) to find the optimal weights whose cost is negligible since it does not involve solving ODEs.
We call this `one subject at a time' (OSAT) method,
which is easily parallelized since the SA runs of each subject are independent.

\section{Results and Discussions}
\label{secResults}
\subsection{Hardware and Software}
\label{secHWSW}
In this paper, we used a ThinkPad P72 laptop with Intel Xeon-2186M CPU (2.9Ghz base frequency and 4.8Ghz max turbo frequency, 6 cores 12 threads) and 64GB DDR4-2666 ECC memory.
NPSA is written in modern Fortran and is fully parallelized using MPI.
We use Intel Fortran and MPI provided in the free Intel OneAPI 2022.1.3, and we run it on Windows 10 Pro.
To achieve the fastest speed of NPAG, we run it using gfortran with OpenMP on Ubuntu.
In both NPSA and NPAG, the absolute tolerance ATOL and relative tolerance RTOL in the ODE solvers are set as $10^{-4}$.

\subsection{Example 1}
\label{secEx1}
Unlike the current nonparametric and semiparametric methods
which have to treat the fixed but unknown parameters $\bm{\beta}$ and the distribution $f$ separately, i.e., find the (local) maximum LL by iterating among fixing $\bm{\beta}$, finding $f$, updating $\bm{\beta}$.
One advantage of NPSA is that there is no need to do such iteration between $\bm{\beta}$ and $f$,
NPSA can stochastically vary $\bm{\beta}$ and $f$ and the same time when exploring the global maximum LL.

In this section,
we pick such a semiparametric (contains $\bm{\beta}$ and $f$) example in Wang's paper \cite{Wang2010semiparametric} which is used to solve overdispersion problem and has a known global maximum LL and several local maximum LL.
Then by comparing with Wang's method,
we show that NPSA can straightforwardly find the global maximum solution.

In this model, the number of subjects $n=20$,
the support point $\bm{\mu}_k$ and $\bm{\beta}$ are both 1 dimensional,
so $p(\bm{Y}_i\vert \bm{\beta}, \bm{\mu}_k)$ can be written as $p(\bm{Y}_i\vert \beta, \mu_k)$ and is given by,
\begin{align}\label{wang2010ex1_model}
p(\bm{Y}_i\vert \beta, \mu_k) =
\left[
p(\mu_k + \beta x_i )
\right]^{y_i}
\left[
1-p(\mu_k + \beta x_i )
\right]^{n_i-y_i},
\end{align}
where $p(\eta)=e^{\eta}/(1+e^{\eta})$.
The data file of $y_i$ and corresponding covariates $n_i$ and $x_i$ can be found exactly in Table 1 in \cite{Wang2010semiparametric}.

\begin{table}[htbp]
\caption{
Comparison among NPSA (using choice 3), Wang's method, and NPAG, using Wang's model and data \cite{Wang2010semiparametric}.
The maximum log likelihood is labeled as "LL",
the optimum solutions for each algorithm are denoted by $\beta$ and the support points (each with label $k$, weight $w_k$ and location $\mu_k$).
The columns of $k$, $w_k$ and $\mu_k$ are sorted according to ascending order with regard to $\mu_k$.
The NPSA runs are coded in Fortran, NPSA-OSAT, NPSA (Choice3) took 0.06 and 1 seconds correspondingly on a single CPU core.
The model and data set for NPAG has not been coded in Pmetrics,
so the NPAG run is coded in MATLAB just to show the results, it took 0.6 seconds.
NPAG uses 1000 grid points.
In NPSA-OSAT, the support points whose weights smaller than $10^{-5}$ are neglected.
}
\label{tab_wangex1compare}
\begin{tabular*} {\textwidth}{@{\extracolsep{\fill}}lcccccccc}
\toprule
& \multicolumn{2}{c}{NPSA (Choice 3)} & \multicolumn{2}{c}{Wang} & \multicolumn{2}{c}{NPAG} & \multicolumn{2}{c}{NPSA-OSAT} \\
& \multicolumn{2}{c}{$\textrm{LL} = -205.42214$ } & 
\multicolumn{2}{c}{$\textrm{LL} = -205.42216$ } &
\multicolumn{2}{c}{$\textrm{LL} = -205.42215$ }  &
\multicolumn{2}{c}{$\textrm{LL} = -205.6253$ } \\
& \multicolumn{2}{c}{$\beta= 0.97007$ } &
\multicolumn{2}{c}{ $\beta=0.970$ } &
\multicolumn{2}{c}{$\beta= 0.97007$\textrm{(fixed)} } &
\multicolumn{2}{c}{$\beta= 0.97007$\textrm{(fixed)} }  \\
$k$ & $w_k$  & $\mu_k$  & $w_k$  & $\mu_k$  & $w_k$  & $\mu_k$  & $w_k$  & $\mu_k$ \\
\cmidrule{2-3}  \cmidrule{4-5}  \cmidrule{6-7} \cmidrule{8-9}
         1 &    0.00201 &   -3.24591  & 0.270 & -3.245  &     0.2677 &    -3.2461  &     0.08942 &    -3.33086 \\

         2 &    0.00706 &   -3.24527  & &  & 1.00E-04 &  -3.2422 & 0.20954 &	-3.2522 \\

         3 &    0.01156 &   -3.24517 \\

         4 &    0.03876 &   -3.24512 \\

         5 &    0.02621 &   -3.24498 \\

         6 &    0.02436 &   -3.24495 \\

         7 &    0.02458 &   -3.24492 \\

         8 &   3.78E-05 &   -3.24491 \\

         9 &    0.13153 &   -3.24487 \\

        10 &    0.00351 &   -3.24389 \\

        11 &    0.05097 &   -2.98143 & 0.130 & -2.981 &     0.1061 &    -2.9836 & 0.10085 &	-2.85204 \\

        12 &     0.0793 &   -2.98122 & & &      0.026 &    -2.9797 \\

        13 &    0.06461 &   -0.70533  & 0.068 & -0.705  &     0.0684 &    -0.7072 & 0.06405 &	-0.80513 \\

        14 &    0.00383 &   -0.70527  \\

        15 &    0.04467 &    0.88589  & 0.532 & 0.886  &     0.2769 &     0.8837 & 0.00957 &	0.45132\\

        16 &    0.02236 &    0.88593 &  &  &     0.2548 &     0.8883  & 0.24195 &	0.57295\\

        17 &    0.19823 &    0.88593 & & & & & 0.28462 &	1.22141 \\

        18 &    0.12567 &    0.88598 \\

        19 &    0.02171 &    0.88601 \\

        20 &    0.11903 &    0.88607 \\
\bottomrule
\end{tabular*}
\end{table}

In Table \ref{tab_wangex1compare},
we list the optimum solution found by NPSA, Wang, and NPAG.
For NPSA, we use choice 3 as mentioned in Sec. \ref{SecAlgo},
and we use $(R_t,N_s,N_t)=(0.5,20,10)$.
For NPAG, we use 1000 uniform random grid points for $\mu_k$ in the initial grid-searching stage.
For both NPSA and NPAG, the range of $\mu_k$ is set between -10 and 10.
The range of the fixed but unknown parameter $\beta$ in NPSA is set between -10 to 10 (Wang's range is between -2 and 4), and $\beta$ found by NPSA and Wang are both around 0.97
(NPAG currently does not search for $\beta$, so its value is fixed).

We found that both of NPSA and NPAG can find slightly higher maximum likelihood (max LL),
than Wang's approximated global optimum LL which is -205.422167,
about 1 second or less on a single CPU core.
NPSA can find even higher max LL than NPAG (-205.422142 .vs. -205.422159).
As shown in Table \ref{tab_wangex1compare},
Wang's method found 4 different support points and corresponding weights,
NPAG found 7 support points with weights,
and NPSA got 20.
However, compared with Wang's method, some of the support points and weights from NPSA and NPAG can be approximated combined.
For example,
the No. 1 through No. 10 support points and their weights from NPSA, and the No.1 and No.2 support points and their weights from NPAG,
can be combined into the No.1 support point from Wang's method with $\mu_k=-3.245$ and $w_k=0.270$.
The No.11 and No.12 support points from NPSA, and the No.3 and No.4 support points from NPAG, can be combined into the No.2 support point from Wang's method.
The No.13 and No.14 support points from NPSA can be combined into the No.3 support points from Wang's method, or the No.5 support points from NPAG.
The No.15 through No.20 support points from NPSA, and the No.6 and No.7 support points from NPAG, can be combined into the No.4 support point from Wang's method.
For all the methods listed in Table \ref{tab_wangex1compare},
the corresponding D functions are all found to be close to zero, indicating global optimal solution is reached.

According to Mallet \cite{Mallet1986a} and Lindsay \cite{Lindsay1983a},
in this model and data,
there will be no more than 20 (the number of subjects) unique support points with corresponding weights.
The results from NPSA, Wang's method, and NPAG, showed that there are approximated 4 unique support points with corresponding weights.
This is indeed an example of the theories of Mallet and Lindsay.

\subsection{Example 2}
\label{secAlan1compEx}

In this section,
we compare NPSA and NPAG by using a classical two-parameter one-compartment PK model.
The two parameters are $K$ and $V$, so
\begin{equation}\label{thetai_ex2}
\bm{\theta}_i = \left( K, V \right).
\end{equation}

The predicted concentration $y_{ji}^{pred}$ for subject $i$ at time $t_j$ is given by
\begin{equation}\label{yjipred_ex2}
y_{ji}^{pred} = \frac{A}{V}e^{-K t_j},
\end{equation}
where the parameter $A$ is fixed at 20.

The observed concentration $y_{ji}$ is given by
\begin{equation}\label{yji_ex2}
y_{ji} = y_{ji}^{pred} + \epsilon_{ji} ,
\end{equation}
where $\epsilon_{ji}$ is the noise term.

For this model,
the noise term $\epsilon_{ji}$ is a gaussian random number whose mean is 0 and standard deviation is $\sigma$.
So the likelihood $p(\bm{\bm{Y}_i\vert\beta}, \bm{\theta}_i)$ for this model is
\begin{align}
p(\bm{\bm{Y}_i\vert\beta}, \bm{\theta}_i) =
p(\bm{Y}_i\vert\sigma^2, \bm{\theta}_i) &=
\prod_{j=1}^{m_i}
\frac{    \exp \left[  -\frac{1}{2} \left(\frac{Y_{ji} - y_{ji}^{pred}  }{\sigma }\right)^2   \right]    }
{ \sqrt{2\pi} \sigma  } ,
\label{Yji_ex2__likelihood}
\end{align}
this assumes measurements are independent for each subject,
where $m_i$ is the total number of time slots for subject $i$.

For the data file,
we use 5 time slots (so $m_i=5$) for each subject, $t=0.2, 0.4, 0.6, 0.8, 1.0$,
we set $\sigma=0.5$ for the noise term.
The $y_{ji}$ is generated from Eq. (\ref{yji_ex2}), in which
$K$ is generated from two mixed Gaussians $N(0.5, 0.05^2)$ and $N(1.5, 0.15^2)$ with equal weights,
$V$ is generated from a Gaussian $N(1.0, 0.2^2)$.
We have tested NPSA-OSAT, NPSA (choice 2), NPSA (choice 3), and NPAG for data files with different number of subjects.
The values of $y_{ji}$ for each subject at each time slot is listed in \cite{2p100} as an example for 100 subjects.

\begin{figure}[!tbhp]
\centering
\includegraphics[width=\columnwidth]{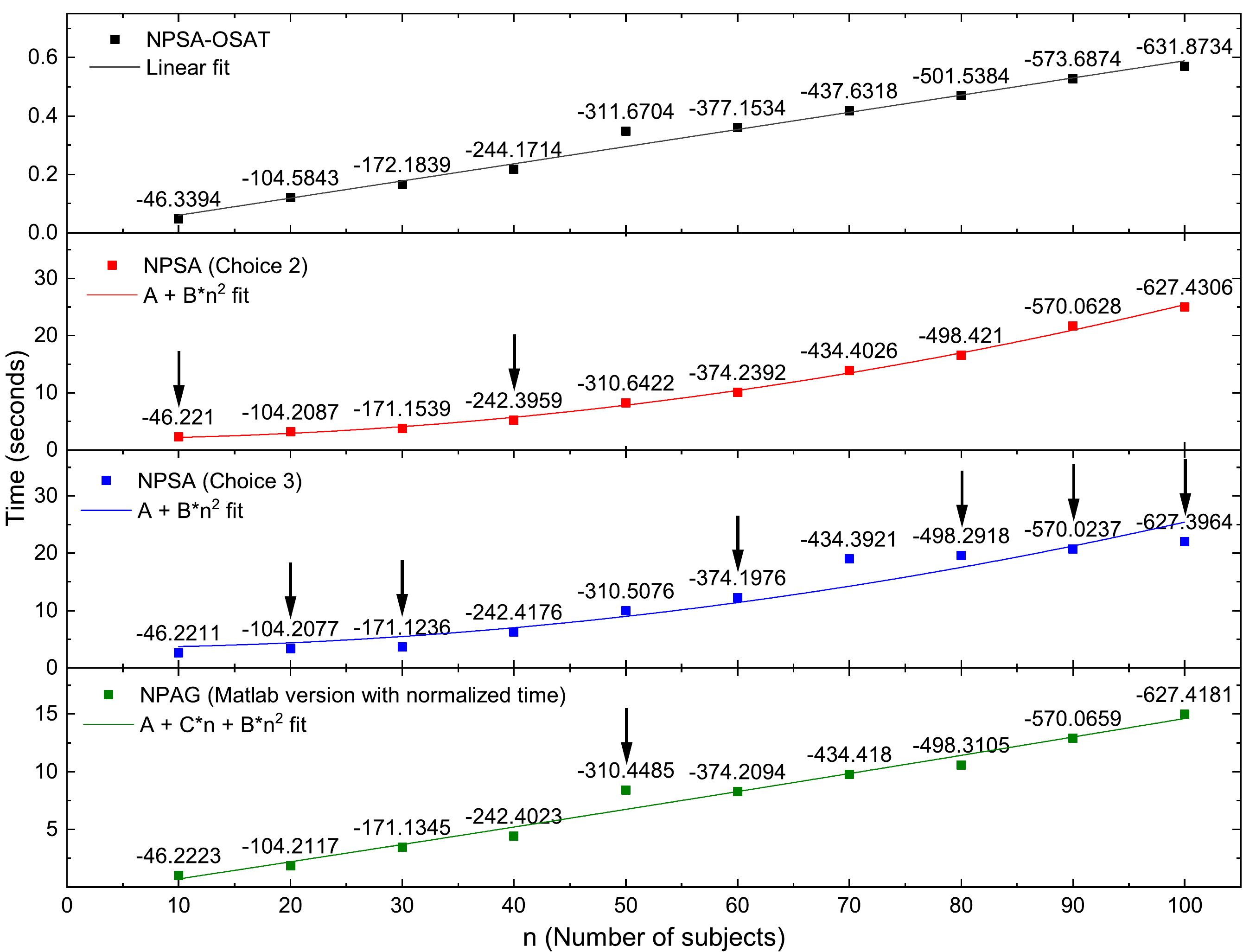}
\caption{
Comparison among NPSA-OSAT, NPSA (Choice2), NPSA (Choice3), and NPAG,
for the 2-parameter one compartment model described in Sec. \ref{secAlan1compEx}.
The 3 NPSA runs are coded in Fortran.
The NPAG run has been coded in MATLAB because this model and data set has not been coded in Pmetrics.
So its running time is not directly comparable with NPSA.
However, the actual running time does not matter too much, since the main purpose is to show how the computation time scales as the number of subjects changes.
Therefore, for NPAG we use its normalized time,
which is the actual running time divided by the time it takes for 10 subjects.
The LL for each run are labeled correspondingly.
The highest LL reached for each given number of subjects is denoted by the down arrow correspondingly.
}
\label{Alan_KV_model_cost}
\end{figure}

\begin{figure}[!tbhp]
\centering
\includegraphics[width=\columnwidth]{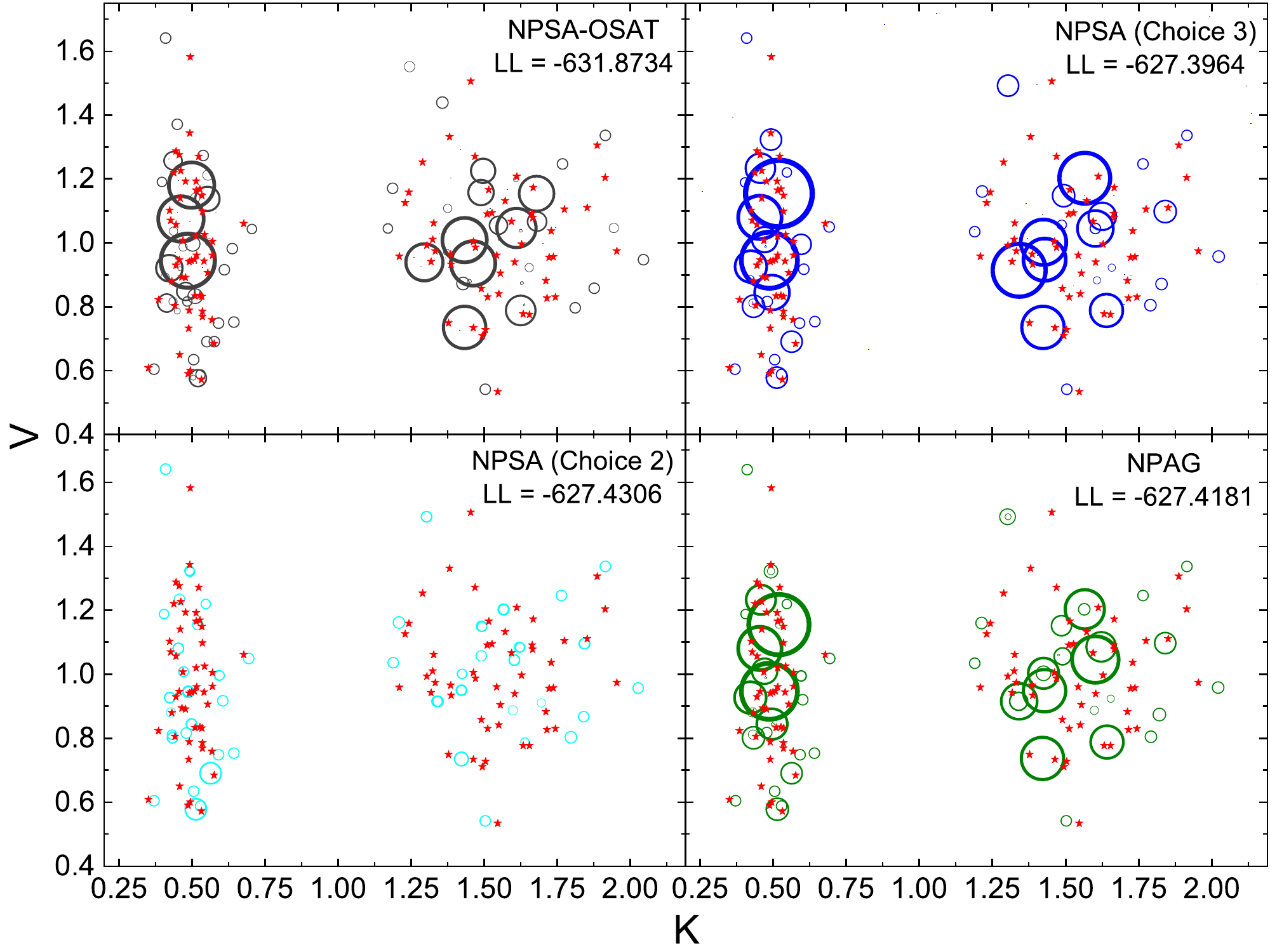}
\caption{
Comparison among NPSA-OSAT, NPSA (Choice2), NPSA (Choice3), and NPAG,
for the 2-parameter one compartment model described in Sec. \ref{secAlan1compEx}.
The maximum LL reached are listed correspondingly as "LL".
In each plot, The support point is in the center of circle, and its weight is indicated by the radius.
The $(K,V)$ pairs used to generate the dataset are presented by red stars.
The 3 NPSA runs are coded in Fortran, NPSA-OSAT, NPSA (Choice2) and NPSA (Choice3) took 0.57, 25, 22 seconds on a single CPU core.
The model and data set for NPAG has not been coded in Pmetrics,
so the NPAG run is coded in MATLAB just to show the results, it took 8 seconds.
}
\label{Alan_KV_model}
\end{figure}

In Fig. \ref{Alan_KV_model_cost},
we list the time cost of NPSA-OSAT, NPSA (Choice2), NPSA (Choice3), and NPAG (with normalized time) for the data sets with different number of subjects $n$.
As discussed in Sec. \ref{secNPSAOSATcost}, as expected, NPSA-OSAT scales linearly with $n$,
therefore it is way much faster than NPSA (Choice2), NPSA (Choice3).
NPSA (Choice2), NPSA (Choice3) both clearly scale with $n^2$ and can be fit well by using $A + B n^2$ where $A$ and $B$ are fitting parameters.
This is expected as discussed in Sec. \ref{secNPSAcost}.
NPAG's cost is $\mathcal{O}(n a^d) + \mathcal{O}(n^2 d)$ as discussed in Sec. \ref{secNPAGcost}.
Since NPAG uses fixed number of grid points in the initial grid search, so $a^d$ is a constant, therefore its cost is $\mathcal{O}(n) + \mathcal{O}(n^2 d)$.
Indeed, we found a quadratic function of $n$ with $A$, $B$, and $C$ as fitting parameters, $A + B n + C n^2$, fits the time cost of NPAG well.
The reason that NPAG looks like it scales linearly is because for this model the grid searching part takes most of the computation time, which means the quadratic term $C n^2$ is much smaller than the linear term $B n$.

About the D function, for all the runs, NPSA-OSAT's D functions are about $3\%$ the magnitude of the corresponding log likelihood (LL).
While for regular NPSA, namely NPSA (Choice2) and NPSA (Choice3),
as well as NPAG, their D functions are about $1\%$ the magnitude of the corresponding LL and their LL are higher than NPSA-OSATs'.
Therefore, NPSA (Choice2), NPSA (Choice3), and NPAG's solutions are likely closer to the global optimum than NPSA-OSATs'.
However, NPSA-OSAT's LL are similar with NPSA and NPAGs', and its D functions are small enough, and what is more, it is way much faster than NPSA and NPAG.
Considering the speed and its ability in achieving relatively high LL, clearly NPSA-OSAT provides an attractive solution.

\subsection{Example 3}
\label{secAlan2compEx}

In this section,
we use a 4-parameter two-compartment model with analytic solution which is similar with the one in Ref. \cite{DArgenio2019analytic}.
The 4 parameters are $K$, $V$, $K_{cp}$, and $K_{pc}$.
So we have,
\begin{equation}\label{thetai_ex2}
\bm{\theta}_i = \left( K, V, K_{cp}, K_{pc}  \right).
\end{equation}

The analytic solution of the predicted concentration $y_{ji}^{pred}$ for subject $i$ at time $t_j$ is given by
\begin{equation}\label{yjipred_ex2}
y_{ji}^{pred} = A e^{- \alpha t_j} + B e^{-\beta t_j} ,
\end{equation}
where
\begin{align}
  A &= \frac{D (\alpha-K_{pc}) }{V (\alpha - \beta) } \\
  B &= \frac{D (K_{pc}-\beta)}{V(\alpha - \beta)} \\
  \alpha &= \frac{ K + K_{cp} + K_{pc} + \sqrt{(K+K_{cp}+K_{pc})^2 - 4 K \times K_{pc} }  }{2} \\
  \beta &= \frac{ K + K_{cp} + K_{pc} - \sqrt{(K+K_{cp}+K_{pc})^2 - 4 K \times K_{pc} }  }{2} \\
  D &= 20 .
\end{align}
This is a more complex model than the one in Sec. \ref{secAlan1compEx}.
In this model, the observed concentration $y_{ji}$ takes the same form as Eq. (\ref{yji_ex2}),
and the noise term $\epsilon_{ji}$ is a gaussian random number whose mean is 0 and standard deviation is $\sigma$.
So, for this model,
the likelihood $p(\bm{\bm{Y}_i\vert\beta}, \bm{\theta}_i)$ takes the same form as Eq. (\ref{Yji_ex2__likelihood}).

For the data file,
we use 5 time slots (so $m_i=5$) for each subject, $t=0.2, 0.4, 0.6, 0.8, 1.0$,
we set $\sigma=0.5$.
The $y_{ji}$ is generated from Eq. (\ref{yji_ex2}), in which
$K$ is generated from two mixed Gaussians $N(0.5, 0.06^2)$ and $N(0.8, 0.06^2)$ with equal weights,
$V$ is generated from a Gaussian $N(1.0, 0.2^2)$,
$K_{cp}$ is generated from a Gaussian $N(0.5, 0.2^2)$,
$K_{pc}$ is generated from a Gaussian $N(2.0, 0.1^2)$.
The data file of 100 subjects can be found in \cite{4p100}.

\begin{figure}[!tbhp]
\centering
\includegraphics[width=\columnwidth]{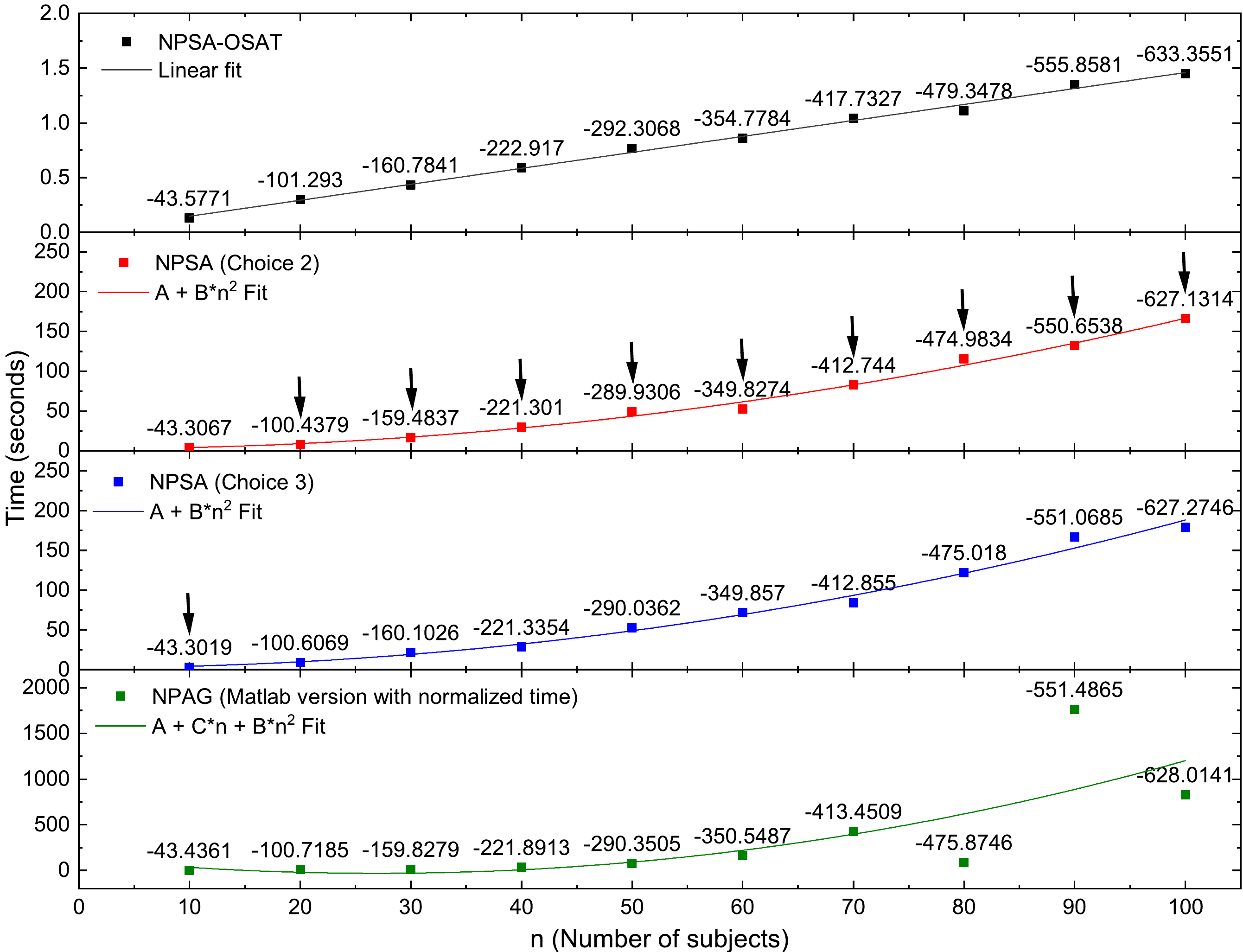}
\caption{
Comparison among NPSA (Choice2), NPSA (Choice3), NPAG, NPSA-OSAT about the computation cost for the two compartment 4-parameter analytic model described in Sec. \ref{secAlan2compEx}.
The 3 NPSA runs are coded in Fortran and are run on a single CPU core.
The NPAG run has been coded in MATLAB because this model and data set has not been coded in Pmetrics.
So its running time is not directly comparable with NPSA.
However, since the main purpose is to show how the computation time scales as the number of subjects changes.
Therefore, similar with what is done in Fig. \ref{Alan_KV_model}, for NPAG we use its normalized time,
which is the actual running time divided by the time it takes for 10 subjects.
The LL for each run is labeled correspondingly.
The highest LL for each given number of subjects is denoted by the down arrow correspondingly.
}
\label{Alan2compEx_cost}
\end{figure}

In Fig. \ref{Alan2compEx_cost},
we list the time cost of NPSA-OSAT, NPSA (Choice2), NPSA (Choice3), and NPAG (with normalized time) for the data sets with different number of subjects $n$.
Again, as expected, NPSA-OSAT scales linearly with $n$, therefore way much faster than regular NPSA and NPAG.
NPSA (Choice2), NPSA (Choice3) both clearly scale with $n^2$.
For a fixed number of grid points in the initial grid search,
NPAG's cost is $\mathcal{O}(n) + \mathcal{O}(n^2 d)$, so a quadratic function of $n$, $A + B n + C n^2$, fits the time cost of NPAG well.

For all the runs,
similar with discussed in Sec. \ref{secAlan1compEx},
all the methods reach about the global optimum solution.
NPSA-OSAT's D functions are about $3\%$ the magnitude of the corresponding LL.
While for regular NPSA and NPAG, their D functions are about $1\%$ the magnitude of the corresponding LL and their LL are slightly higher than NPSA-OSATs'.
However, considering the fast speed and its ability in achieving relatively high LL,
NPSA-OSAT undoubtedly provides an attractive solution again.

\subsection{Example 4}
\label{SecExampleVori}
\subsubsection{Voriconazole model and data}
\label{secModelDate}

For relatively simple models with analytic solutions, both NPAG and NPSA can perform well.
As the model and data become more and more challenging,
we expect to see NPAG and NPSA behave differently.

In this section,
we use the model and the data file which are the same as in those described in detail in our novel Monte Carlo parametric expectation maximization algorithm paper
`RPEM: Randomized Monte Carlo Parametric Expectation Maximization Algorithm' \cite{RPEM2022}.
We use a Voriconazole model \cite{neely2012Pmetrics,Neely2015a,yamada2020npag}
and we follow the data and model format for Pmetrics \cite{yamada2020npag}.
The 7 primary parameters are $K_a$, $V_{max0}$, $K_m$, $V_{c0}$, $F_{A1}$, $K_{cp}$, and $K_{pc}$.
So we have,
\begin{equation}\label{thetaiode}
\bm{\theta}_i = \left(K_a,V_{max0},K_m,V_{c0},F_{A1},K_{cp},K_{pc}\right).
\end{equation}
The covariate is weight ($wt$).
The secondary parameters which are obtained from primary parameters and covariate are $V_m$ and $V$,
\begin{align}
V_m &= V_{max0} \times {wt}^{0.75},
\\
V &= V_{c0} \times {wt}.
\end{align}

For any subject $i$, the ODEs are listed as below,
\begin{align}
\frac{d x_1}{dt} &= -K_a \times x_1,
\\
\frac{d x_2}{dt} &= -K_a \times x_1
+ r^{(i)}_{\rm{IV}}(t)
- \frac{V_m^{(i)} \times x_2}{K_m \times V^{(i)} + x_2}
-K_{cp} \times x_2
+K_{pc} \times x_3,
\\
\frac{d x_3}{dt} &= K_{cp} \times x_2
-K_{pc} \times x_3,
\end{align}
where for subject $i$, $r^{(i)}_{\rm{IV}}(t)$ is the ratio between dose and duration at time $t$. If at time $t$ the dose is non-zero and duration is zero, it means a bolus and $x_1(t)$ needs to be added by an additional $\textrm{dose} \times F_{A1}$.
$V_m^{(i)}$ and $V^{(i)}$ are its secondary parameters $V_m$ and $V$.

In this model, the observed concentration $y_{ji}$  for subject $i$ at time $t_j$ takes the same form as Eq. (\ref{yji_ex2}) and we assume the noise $\epsilon_{ji}$ is a Gaussian random number whose standard deviation $\sigma_{ji}$ has the following form,
\begin{align}
\sigma_{ji} = c_0
+ c_1 \times y_{ji}^{pred},
\label{sigmaji}
\end{align}
where $c_0=0.02$, $c_1=0.1$ are set for this model. The prediction $y_{ji}^{pred}$ is given by
\begin{equation}
y_{ji}^{pred}(\bm{\theta}_i) = \frac{x_2(t_j)}{V^{(i)}}.
\label{yjipred}
\end{equation}

By using Eq.(\ref{stage1}) and Eq.(\ref{stage1_sigma2}),
for this model we have
$
\bm{h}_i(\bm{\theta}_i) = \left[ y_{1i}^{pred},  y_{2i}^{pred}, \ldots,  y_{m_i i}^{pred}    \right]
$
and
$
\bm{G}_i(\bm{\beta}, \bm{\theta}_i) = \text{diag} \left( \sigma^2_{1i}, \ldots, \sigma^2_{m_ii}   \right)
$.
Therefore
\begin{align}
p(\bm{\bm{Y}_i\vert\beta}, \bm{\theta}_i) =
p(\bm{Y}_i\vert\sigma_{ji}^2, \bm{\theta}_i) &=
\prod_{j=1}^{m_i}
\frac{    \exp \left\{  -\frac{1}{2} \left[\frac{Y_{ji} - y_{ji}^{pred}(\bm{\theta}_i)  }{\sigma_{ji} }\right]^2   \right\}      }
{ \sqrt{2\pi} \sigma_{ji}  } .
\label{Yjiode}
\end{align}

\begin{figure}[!hptb]
\centering
\includegraphics[width=\columnwidth]{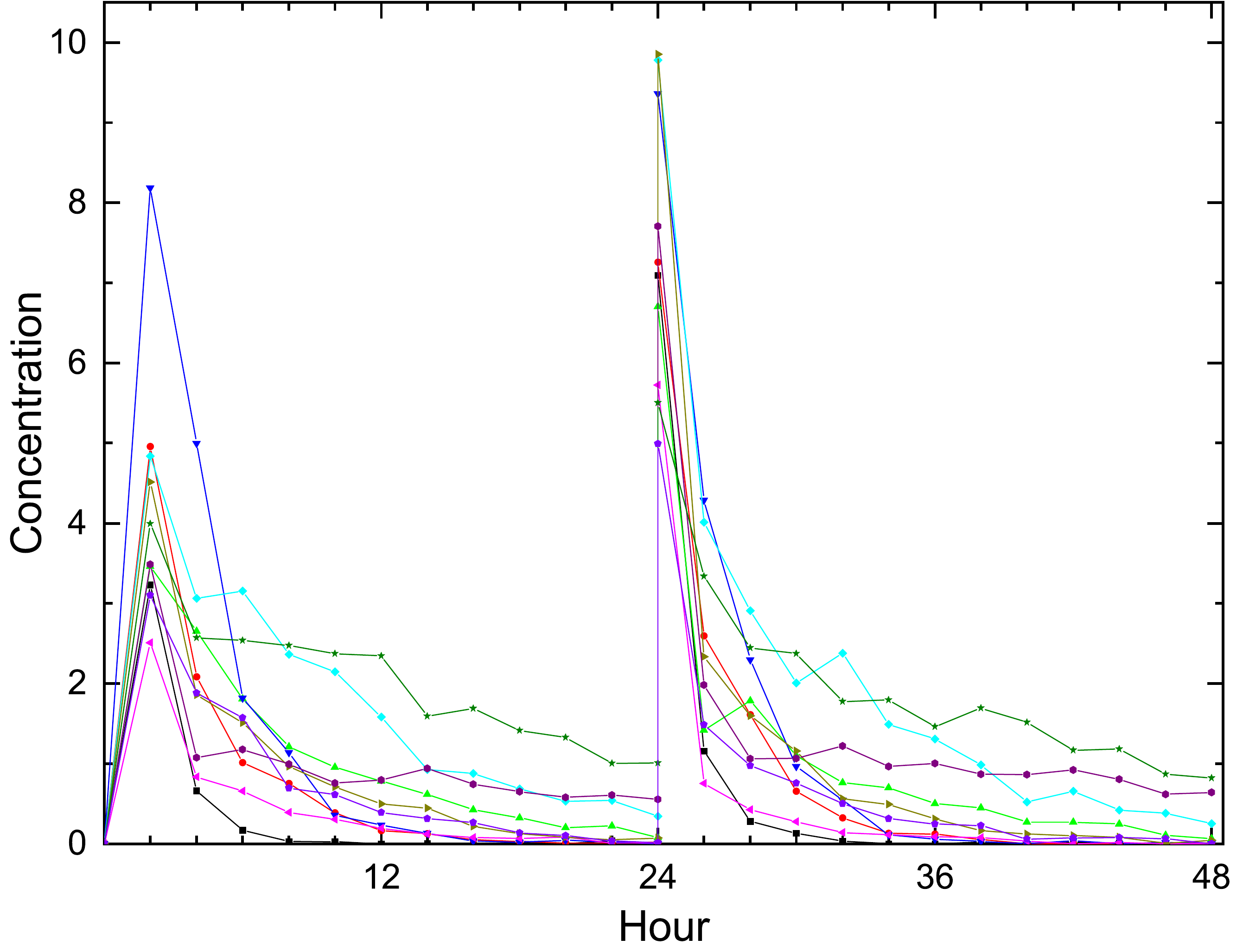}
\caption{Illustration of the simulated concentration vs. time. 10 randomly selected subjects are used. }
\label{Vori_showall}
\end{figure}

The simulated data file for the Voriconazole ODE model can be found in \cite{Voridatafix}.
The observation concentration data $y_{ji}$ is provided as the ``OUT'' column.
They can be generated by using the 50 sets of primary parameters in \cite{Vorisimparfix}.
We randomly select 10 subjects and illustrate the simulated concentration vs. time in Fig. \ref{Vori_showall}.

\subsubsection{Comparison between NPSA and NPAG}
\label{secComparison}

For the current NPSA and NPAG,
the objective functions to be maximized are set as the  log likelihood.
In this section, we compare them in term of the speed and the ability in locating the maximum  log likelihood, as well as the population and individual predictions.

\begin{figure}[!hptb]
\centering
\includegraphics[width=\columnwidth]{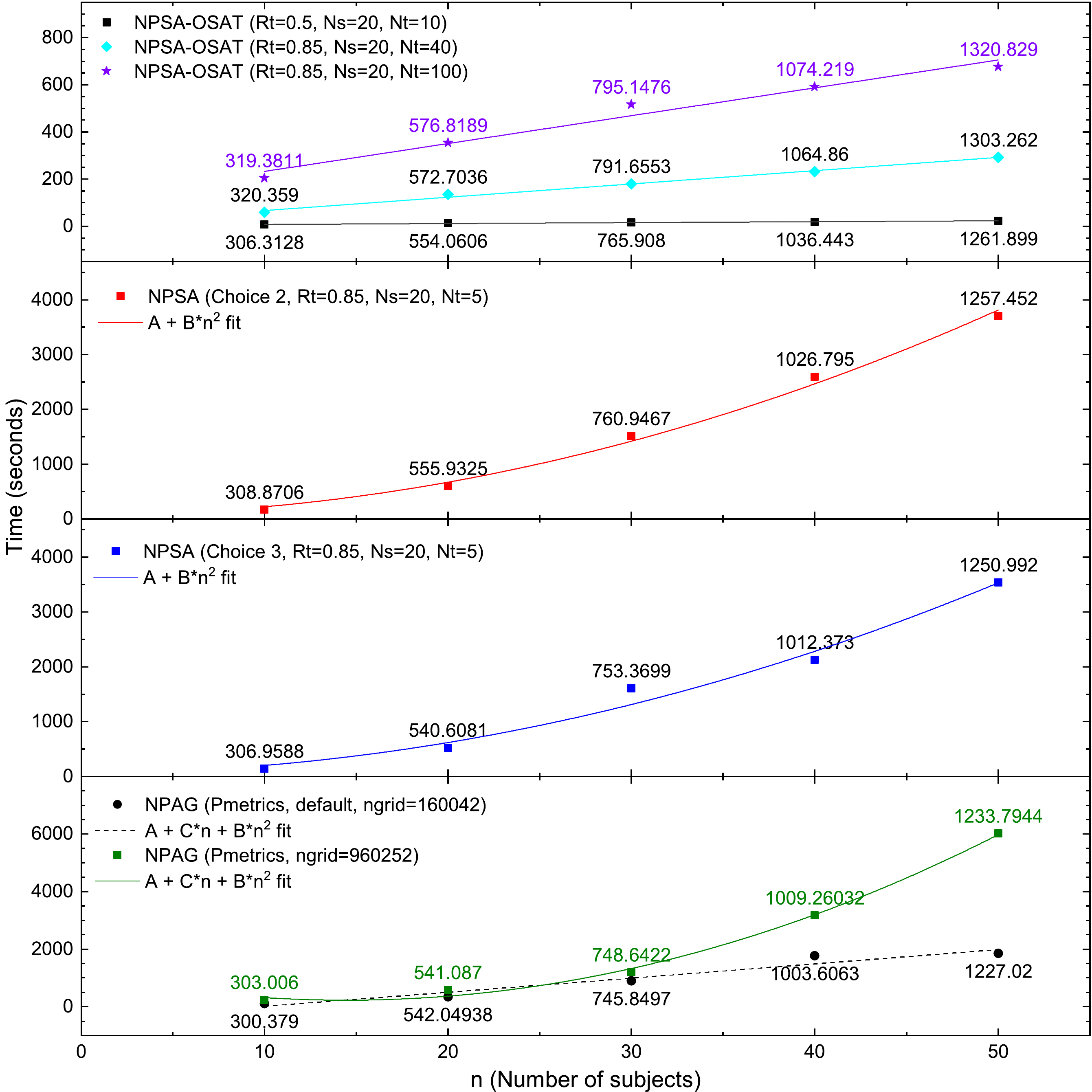}
\caption{
Time cost vs. number of subjects. NPAG and NPSA comparisons.
The labels are  LL.
}
\label{Vori_cost_fit}
\end{figure}

\begin{figure}[!hptb]
\centering
\includegraphics[width=\columnwidth]{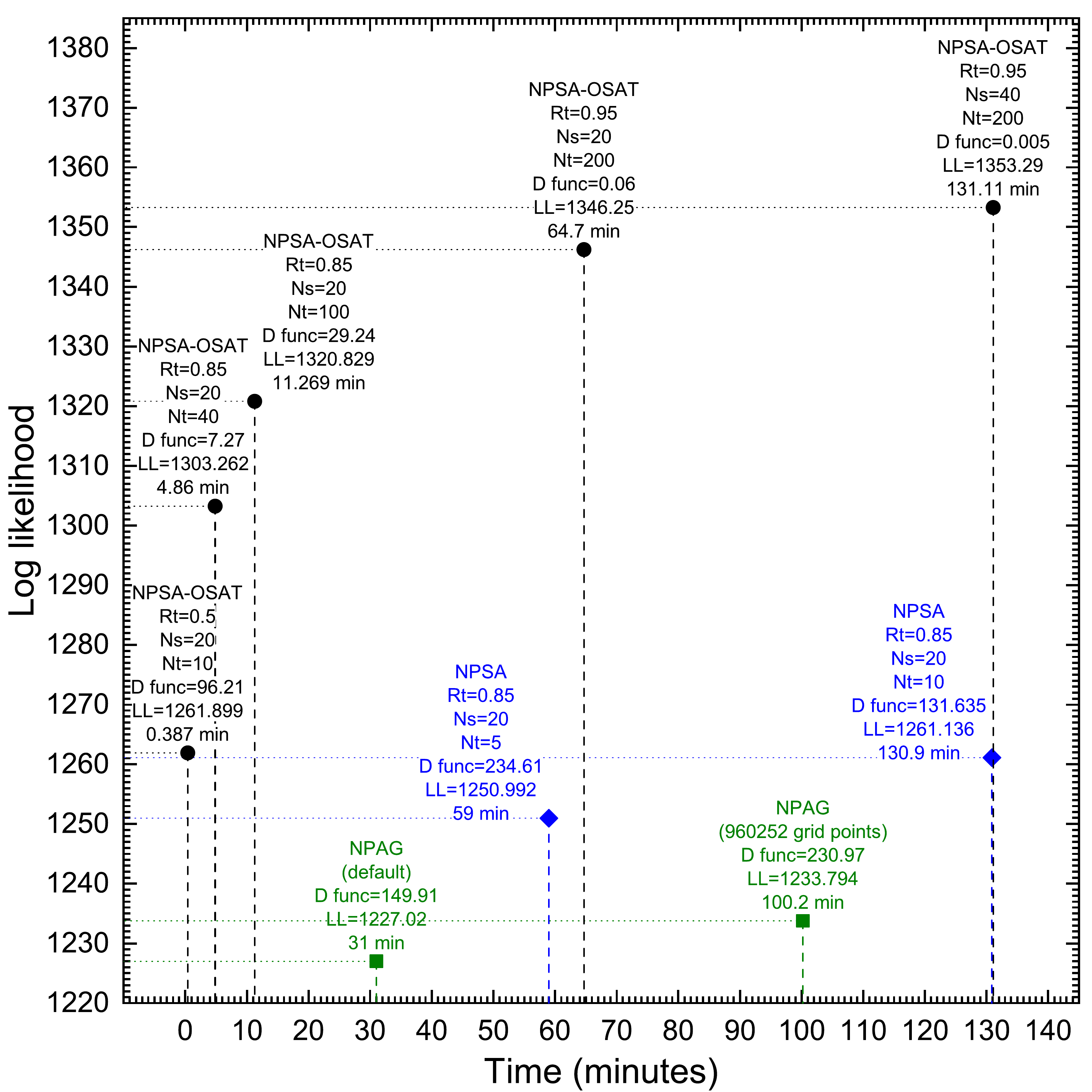}
\caption{
Log likelihood achieved vs. time cost in minutes for NPAG, NPSA (which is NPSA choice 3), and NPSA-OSAT.
The number of subjects are 50.
The value of D function in Eq.(\ref{Dfunc}) are labeled as ``D func''.
}
\label{Vori_cost_selected}
\end{figure}

Similar with Fig. \ref{Alan_KV_model_cost} and Fig. \ref{Alan2compEx_cost}
in Sec. \ref{secAlan1compEx} and Sec. \ref{secAlan2compEx},
in Fig. \ref{Vori_cost_fit},
we show the computational cost of NPAG, NPSA, and NPSA-OSAT with regard to the number of subjects $n$.
As expected,
the computational cost of NPSA-OSAT scales linearly with $n$,
NPSA (choice 2 and 3) scales quadratic with $n$,
and for a fixed number of grid points, NPAG scales as second-degree polynomial function of $n$.

As to the ability in locating the maximum LL, take $n=50$ for example,
for NPAG,
the final LL it can reach heavily depends on the number of initial grid points it searches,
which, as expected, suffers from the curse of dimensionality.
We can see this from the bottom panel in Fig. \ref{Vori_cost_fit}.
Even if we increase the number of initial grid points from the 160042 (NPAG's default)
to 960252 (the maximum grid points can be used for NPAG),
the corresponding total computation time increased from about 30 minutes to about 90 minutes,
NPAG's final LL only increased from 1227.02 to 1233.7944.
Such an increase is less than $1\%$ therefore negligible.
In contrast,
NPSA choice 2 and choice 3 can achieve more than 1250 LL in about 66 minutes,
NPSA-OSAT can reach LL 1261.899 within 25 seconds, and 1320.829 within 12 minutes.
We can conclude that, for this Voriconazole model,
NPSA-OSAT clearly outperforms NPAG and the two regular NPSA in terms of speed and the ability in locating the maximum LL.

In Fig. \ref{Vori_cost_selected},
for 50 subjects,
the LL vs. time (in the unit of minute) for NPSA-OSAT,
NPSA choice 3 (since NPSA choice 2 performs similarly with the default NPSA choice 3, we only include the results from NPSA choice 3 for a cleaner looking of the figure, and we label NPSA choice 3 simply as NPSA), and NPAG with different settings
and the corresponding D functions are presented.
Clearly,
NPSA-OSAT with $(R_T, N_s, N_t) = (0.5, 20, 10), (0.85, 20, 40), (0.85, 20, 100)$ get noticeably higher  LL within much shorter time (within 0.4 to 11 minutes) than NPAG and NPSA.
The NPSA took comparable computation time with NPAG, and reach higher LL than NPAG.
To further check NPSA-OSAT,
we increase the total computational time by increasing $R_T, N_s, N_t$,
so that the temperature in SA cooled down more slowly,
and SA explored the parameter space more thoroughly.
We see that for
$(R_T, N_s, N_t) = (0.95, 20, 10), (0.95, 40, 100)$,
the LL reached 1346.25 and 1353.29, and D functions are 0.06 and 0.005 respectively.
These small D functions indicated that we are around the global optimum solution.
Furthermore, from the results of NPSA-OSAT, the tendency of LL vs. time indicated that the LL probably saturated at around 1353 or so, and further increasing computation time will unlikely to increase the  LL noticeably.
Therefore, taken both the small D functions and the saturated tendency of LL vs. time into consideration,
we can conclude that the solution with LL equals 1353.29 is our global optimum solution for the Voriconazole model and data.

However, we need to point out that,
although NPSA-OSAT is much faster than NPSA and NPAG,
in general, it is not always guaranteed to find the true global maximum LL.
Here for this Voriconazole model and data,
it turned out to be that NPSA-OSAT converges to global maximum LL way much faster than NPSA and NPAG.
In principle, given enough computation time, NPSA should achieve higher LL than NPSA-OSAT, because NPSA is the true global optimizer for the  LL.

\begin{figure}[!hptb]
\centering
\includegraphics[width=\columnwidth]{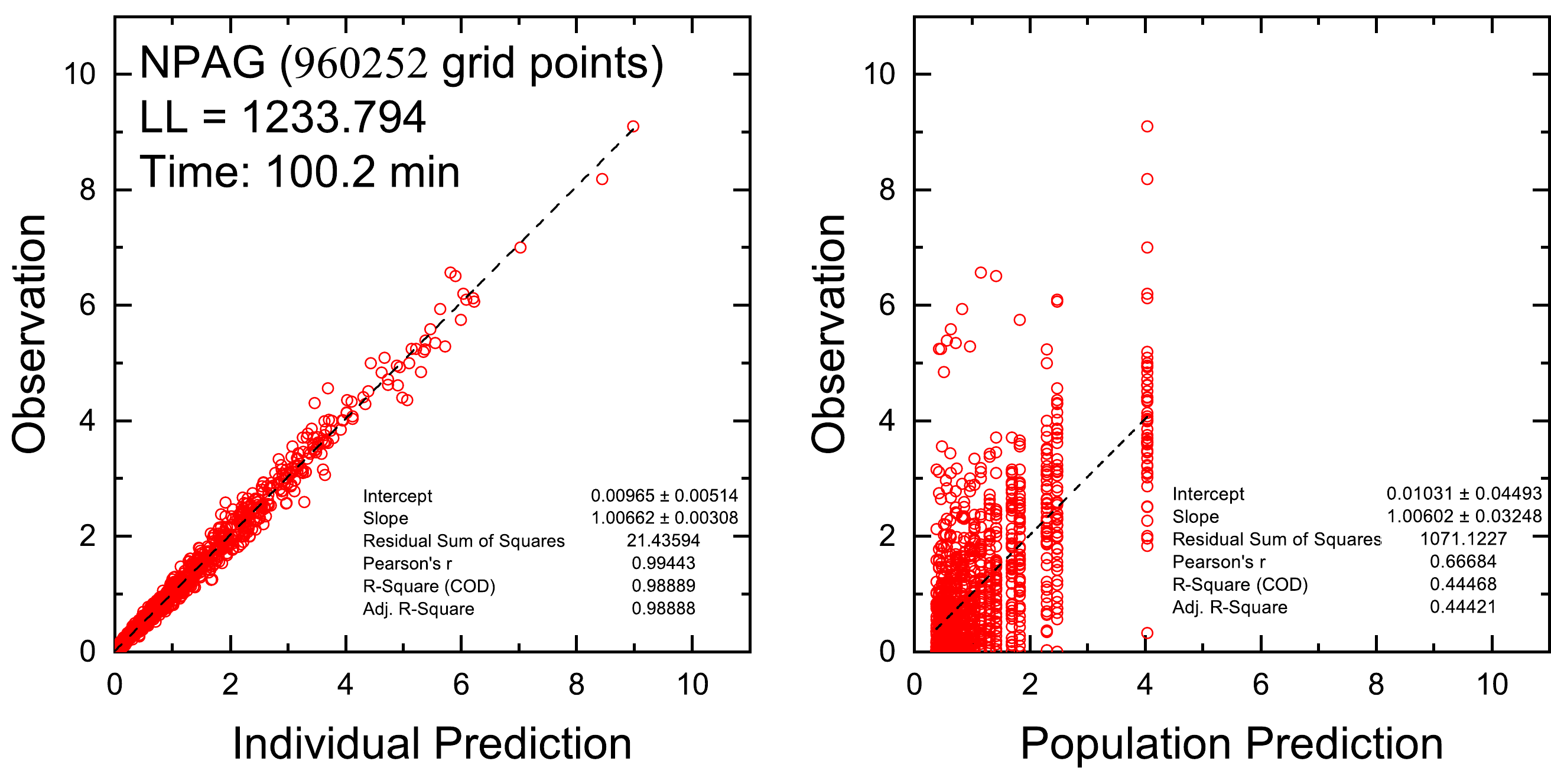}
\caption{
The observation vs. individual prediction,
and the observation vs. population prediction
for NPAG with 960252 initial grid points which is maximum number of grid points the current Pmetrics supports.
It took 100.2 minutes and reached a maximum LL of 1233.794,
and its D function is 230.97.
}
\label{NPAG1233_pred}
\end{figure}

\begin{figure}[!hptb]
\centering
\includegraphics[width=\columnwidth]{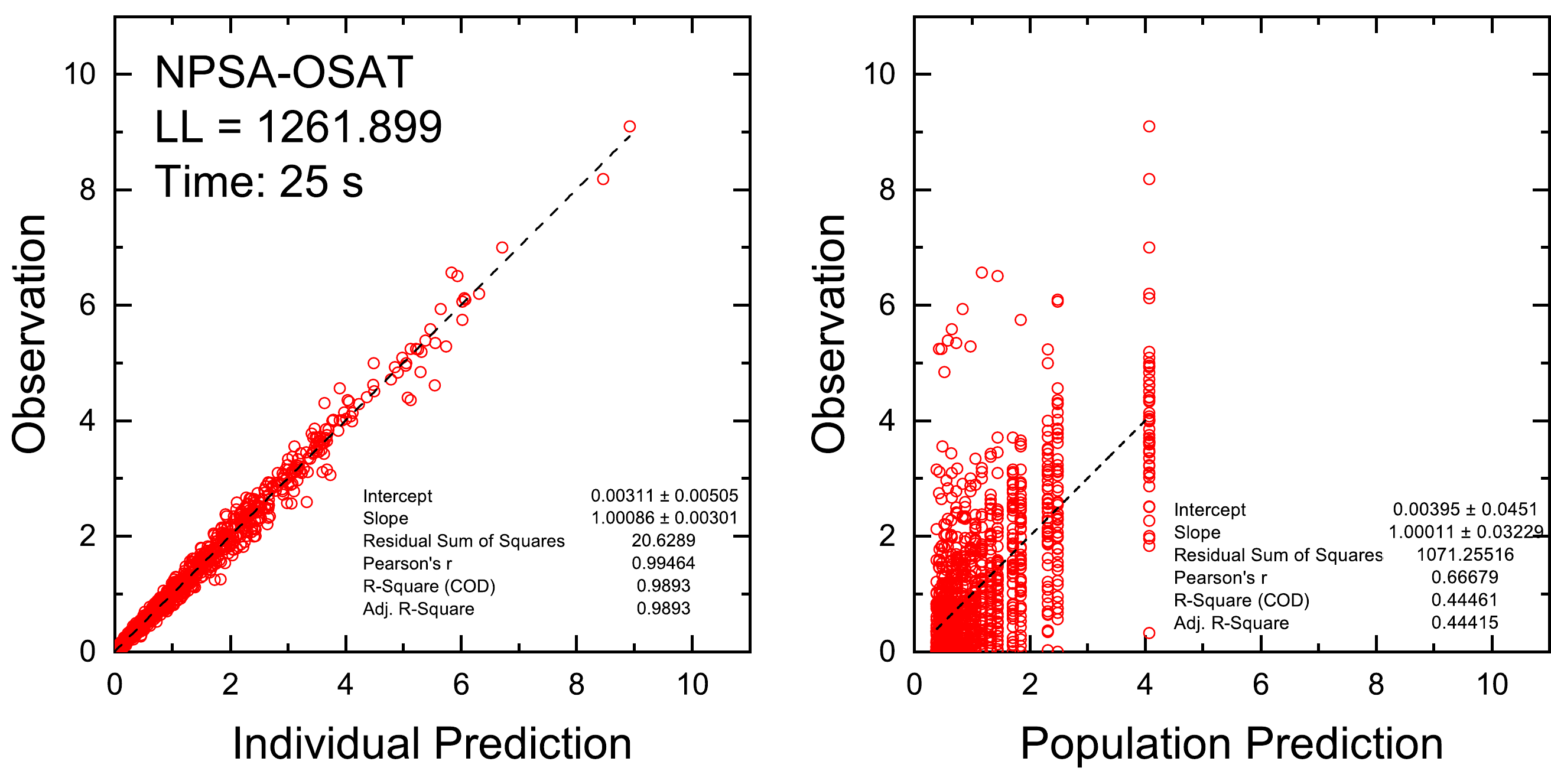}
\caption{
The observation vs. individual prediction,
and the observation vs. population prediction
for NPSA-OSAT with $(R_T=0.5, N_s=20, N_t=10)$.
It took 25 seconds and reached a maximum LL of 1261.899, and its D function is 96.21.
}
\label{NPSAOSAT1261_pred}
\end{figure}

\begin{figure}[!hptb]
\centering
\includegraphics[width=\columnwidth]{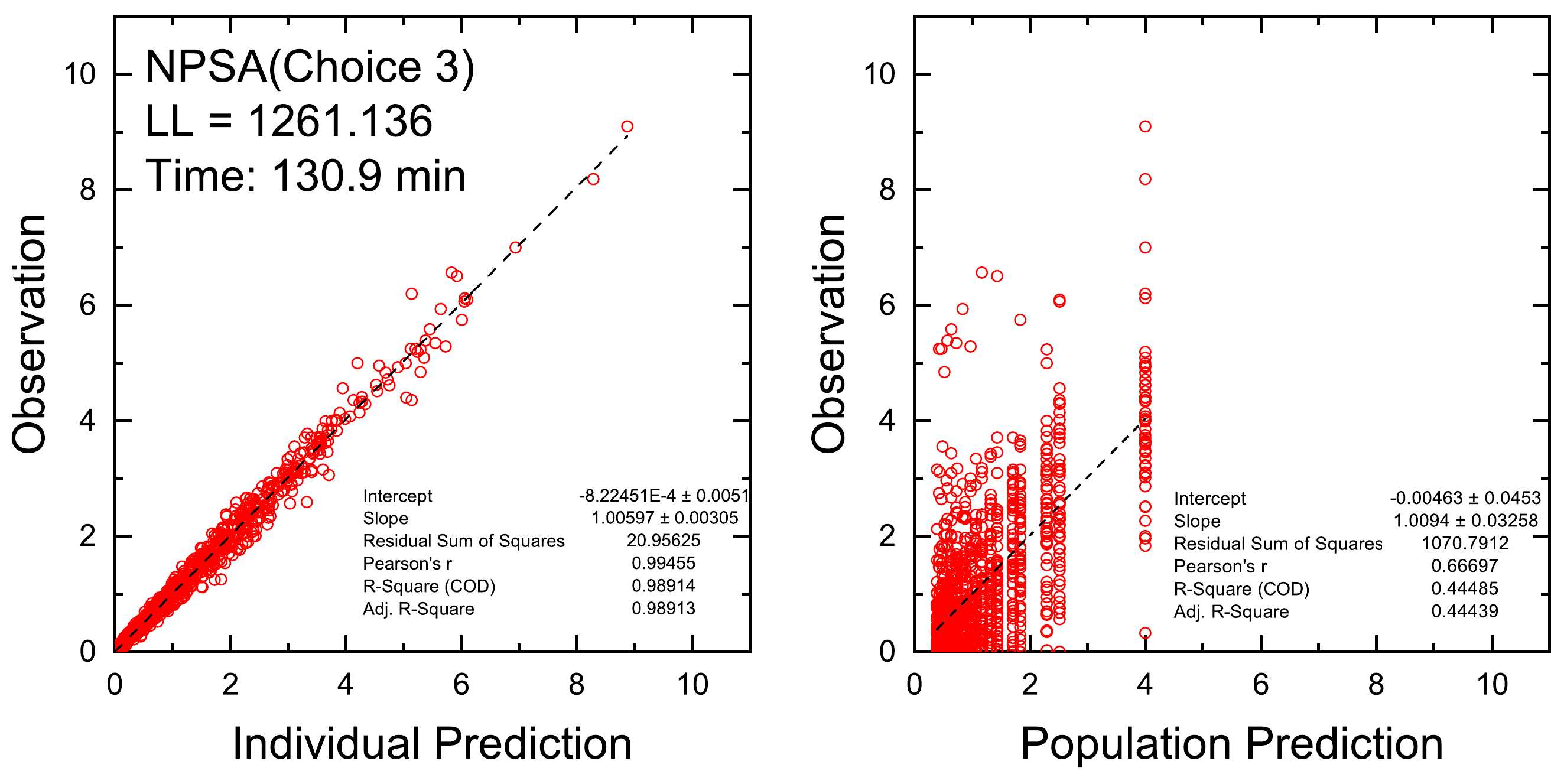}
\caption{
The observation vs. individual prediction,
and the observation vs. population prediction
for NPSA (choice 3) with $(R_T=0.85, N_s=20, N_t=10)$.
It took 130.9 minutes and reached a maximum LL of 1261.136,
and its D function is 131.635.
}
\label{NPSA3_1261_pred}
\end{figure}

\begin{figure}[!hptb]
\centering
\includegraphics[width=\columnwidth]{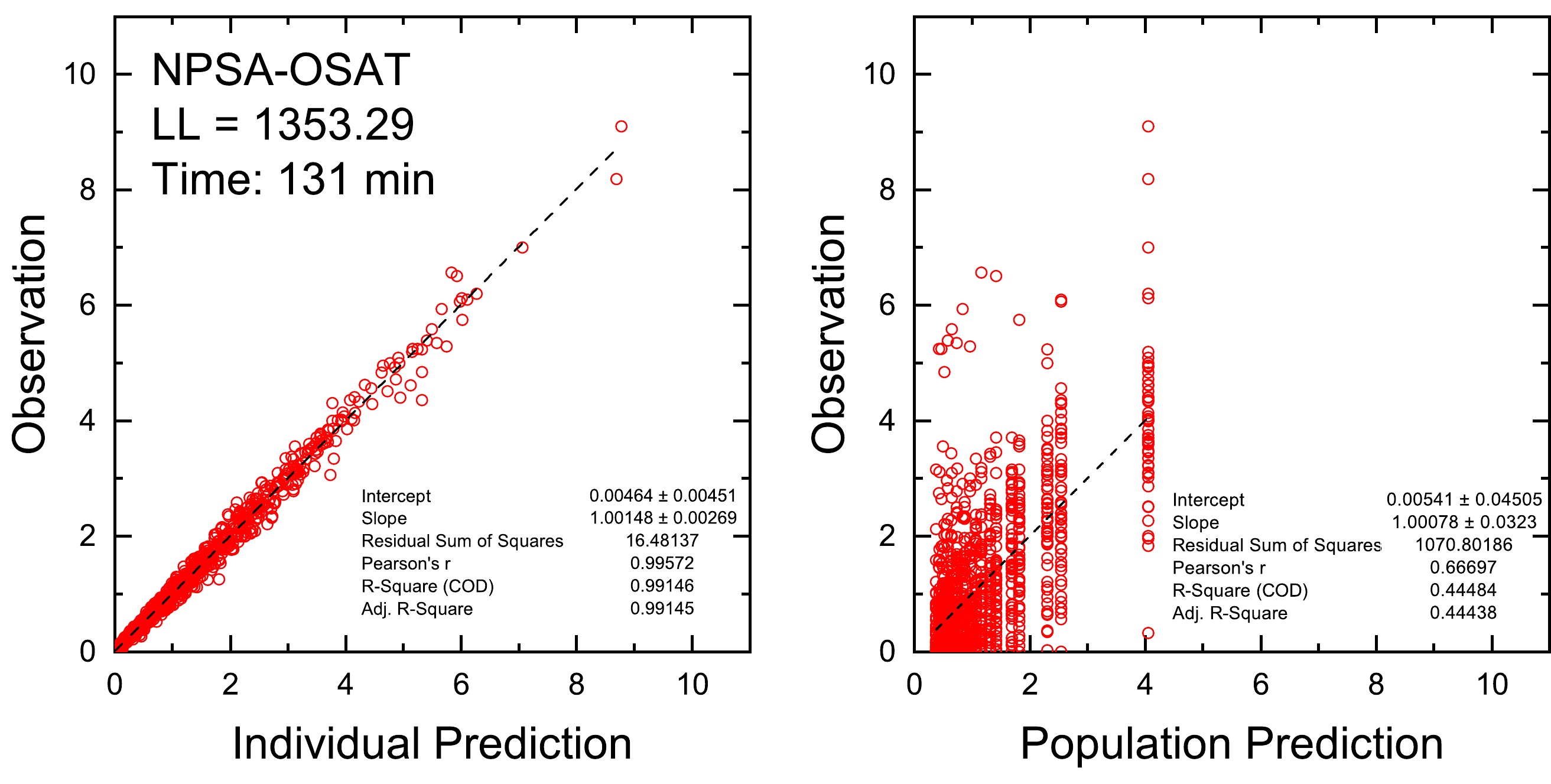}
\caption{
The observation vs. individual prediction,
and the observation vs. population prediction
for NPSA-OSAT with $(R_T=0.95, N_s=40, N_t=200)$.
It took 131.11 minutes and reached a maximum LL of 1353.29,
and its D function is 0.005.
}
\label{NPSAOSAT1353_pred}
\end{figure}

Besides LL, we also care about the predictions.
As to predictions, for subject $i$ at time $t_j$, the population prediction $\langle y_{ji}^{pred} \rangle_{pop} $
is the expectation of $y_{ji}^{pred}$ in Eq.(\ref{yjipred}) using $f(\bm{\theta}_i)$ which is the normalized population distribution (prior) in Eq.(\ref{nonparametric_p}).
$\langle y_{ji}^{pred} \rangle_{pop} $ can be written as,
\begin{align}
\langle y_{ji}^{pred} \rangle_{pop}
= \int y_{ji}^{pred}(\bm{\theta}_i) f(\bm{\theta}_i) d \bm{\theta}_i
= \sum_{k=1}^K w_k y_{ji}^{pred}(\bm{\mu}_k) ,
\label{Yji_pop_pred}
\end{align}
where $w_k$ and $\bm{\mu}_k$ are the estimated weights and support points.

For individual prediction $\langle y_{ji}^{pred} \rangle_{ind}$,
it is the expectation of $y_{ji}^{pred}$ in Eq.(\ref{yjipred}) using the normalized posterior distribution, $\langle y_{ji}^{pred} \rangle_{ind}$ can be written as,
\begin{align}
\langle y_{ji}^{pred} \rangle_{ind}
=
\frac{ \int y_{ji}^{pred}(\bm{\theta}_i) p(\bm{Y}_i\vert {\bm{\beta},\bm{\theta}}_i) f(\bm{\theta}_i) d \bm{\theta}_i  }
{ \int p(\bm{Y}_i\vert {\bm{\beta},\bm{\theta}}_i)  f(\bm{\theta}_i) d \bm{\theta}_i }
=
\frac{ \sum_{k=1}^K y_{ji}^{pred}(\bm{\mu}_k) w_k n_{ik} }{N_i},
\label{Yji_ind_pred}
\end{align}
where $N_i$ and $n_{ik}$ are given by Eq.(\ref{Ni}) and Eq.(\ref{nonparametric_n_ik}).

From figures \ref{NPAG1233_pred} to \ref{NPSAOSAT1353_pred},
based on the results in Fig. \ref{Vori_cost_selected},
we present the population predictions and individual predictions from some of the runs of NPSA-OSAT, NPSA (choice 3), and NPAG.
The reason for the distinct vertical patterns from figures \ref{NPAG1233_pred} to \ref{NPSAOSAT1353_pred} for all the observation vs. population prediction,
is because we set the covariates the same for all the subjects, which lead to the same population predictions for all the subjects.
As to the observation vs. individual predictions, for all the runs the R-square score are all over 0.98 and the slopes of linear fits are all very close to 1,
which means for all the runs, the predictions are similar with the observations,
so they all give reasonably good results.
It is worthwhile to note that, NPSA-OSAT is able to reach LL=1261.899 in 25 seconds,
it achieves higher LL and is at least an order of magnitude faster than NPAG and regular NPSA due to its computational cost is $\mathcal{O}(n)$ instead of $\mathcal{O}(n^2)$,
as pointed out in Sec. \ref{secNPSAOSATcost}.
We also found that, as the LL increases, the linear fit of observation vs. individual predictions becomes better,
which can be seen from the residual sum of squares of the linear fit.
For example, in Fig. \ref{NPSAOSAT1353_pred}, for the NPSA-OSAT run with the highest LL which is 1353, its sum of squares is about 16.48, while for all the rest NPAG and NPSA runs, their sum of squares are about 20 to 21.
Therefore, a high LL does improve the individual predictions.

\subsubsection{Scalability of NPSA for Voriconazole model and data}

NPSA is written in high performance modern Fortran and equipped with MPI \cite{curcic2021toward,sof2022}.
We ran NPSA on the Agave supercomputer cluster at Arizona State University (ASU) and tested its parallel scalability for the Voriconazole model and data.

\begin{figure}[!hptb]
\centering
\includegraphics[width=\columnwidth]{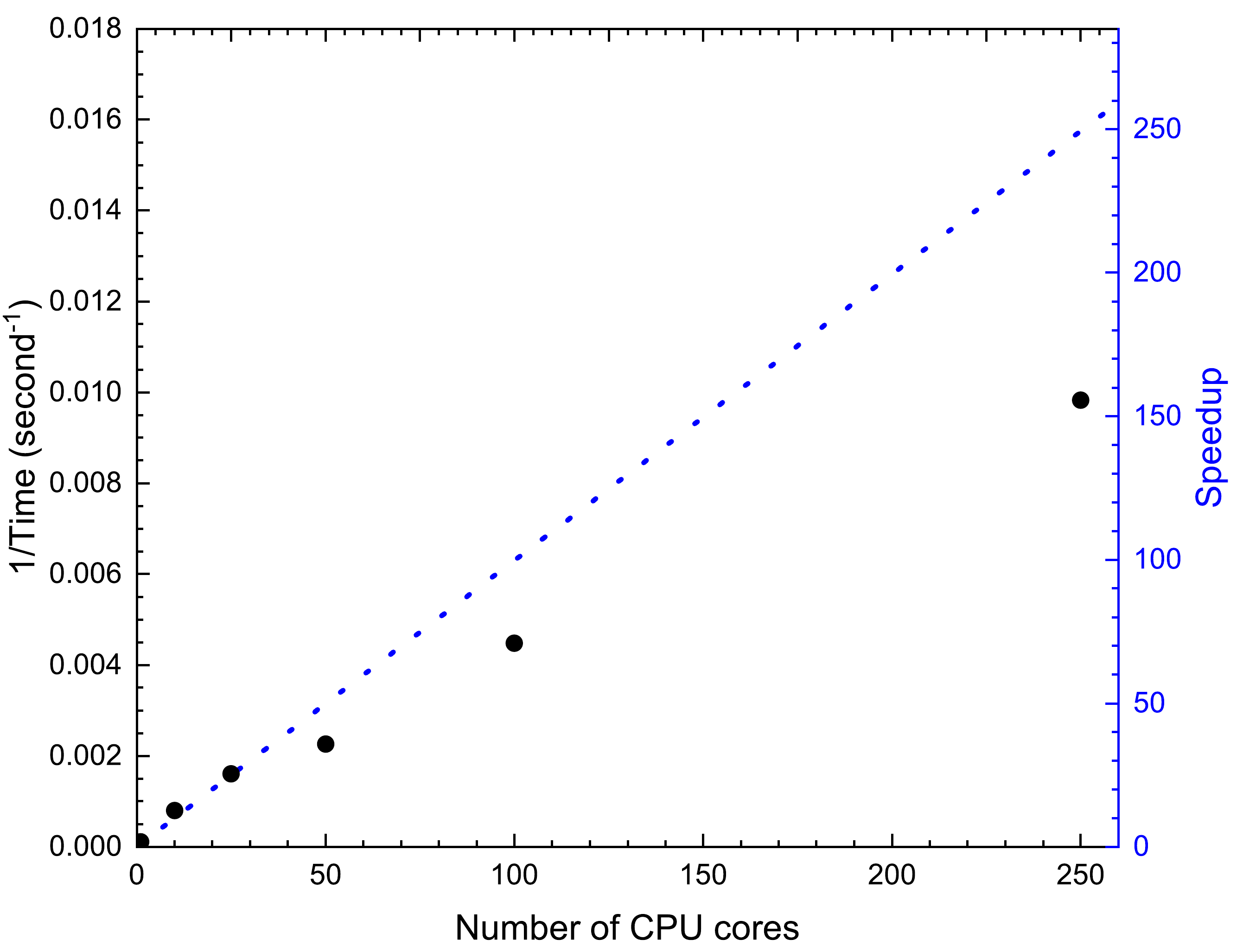}
\caption{The parallel efficiency of the current NPSA code on Agave supercomputer cluster at Arizona State University, for the Voriconazole model and data.
The blue dotted line is the theoretical speedup. The back dots are actual speedup for the corresponding number of CPU cores used.
We use choice 3 (choice 2 of NPSA have similar parallel efficiency),
and the NPSA parameters are $R_t=0.7$, $N_s=20$, $N_t=5$, the initial temperature is set as 60.}
\label{speedup}
\end{figure}

In Fig. \ref{speedup}, we show the scaling efficiency of NPSA.
Overall, when the number of CPU cores are less than 100 (which covers the range from a laptop to high-end personal desktop), the efficiency of NPSA is above $70\%$.
As the number of cores exceeds 100, the ODEs distributed to each CPU cores become fewer and fewer, so the computation load becomes lower and lower for each CPU core,
and the MPI communication time becomes dominant, i.e., the master core distributes jobs to each CPU core to calculate Eq.(\ref{psi_mat}),
collects the results from each CPU core back and then do Metropolis judgement.
Nonetheless, NPSA is still able to maintain above $60\%$ efficiency when the number of CPU cores is 250.
Beyond 250 CPU cores, there is almost no performance gain
by using more CPU cores for this model and data,
since the cost is inevitably dominated by the MPI communications,
whose speed depends on the network used in ASU's Agave cluster.
The time cost of the Voriconazole model and data saturated at around 100 seconds, no matter how many CPU cores are used.

For more complicated models with more subjects, as long as the ratio between computation time and the MPI communication time is high, NPSA will always achieve reasonably high parallel efficiency.

\section{Conclusions}
\label{secSummary}

In this paper,
we discuss our NPSA algorithm,
which is the first application of simulated annealing in nonparametric global optimization.
From analytic models to a Voriconazole model given by ODEs,
we compared NPAG and NPSA under different parameter settings.
The properties of the current NPSA can be summarized as follows.

First,
NPSA is a global optimization algorithm,
which means it is less sensitive than NPAG to the initial conditions, and can find higher LL than NPAG.
This is an important feature because NPSA, unlike NPAG,
does not need to start from many different initial conditions in order to find the best  solution among those local optimal solutions.
Instead, we only need to run NPSA just once and achieve better results than NPAG.
As is shown from the examples in Sec. \ref{secEx1} to Sec. \ref{SecExampleVori},
from simple analytic model to a realistic Voriconazole model, NPSA can almost always achieve higher LL than NPAG and finish within reasonable computation time.

Second, as is shown in Sec. \ref{secCost},
NPSA, as a Monte Carlo method, does not suffer from curse of dimensionality.
Unlike the cost of NPAG which has $\mathcal{O}(n a^d)$ dependence,
NPSA's cost is simply $\mathcal{O}(n^2 d N_t N_s)$ which does not depend on the dimension $d$ exponentially.
Because of this, as the dimension and complexity of the problem increases,
we believe NPSA is likely outperform NPAG or any other grid-searching based methods \cite{yamada2020npag}
not only in the ability of finding the global optimum solution
but also in computational cost.

Third, unlike the conventional way,
NPSA is essentially a "semiparametric" method, in the sense that
NPSA can find the fixed but unknown parameters such as $\beta$ in $\phi$ in Eq.(\ref{nonparametric_phi}) stochastically, since it treats all the parameters in $\phi$ in Eq.(\ref{nonparametric_phi}) equally.
As is shown in Sec. \ref{SecAlgo}.
This feature makes it more desirable than way NPAG finds the value for fixed but unknown parameters.
The NPAG method has to fix $\beta$, then solve the nonparametric problem,
update $\beta$,
then re-solve the NP problem,
then fix $\beta$,
and continue to iterate the process until convergence.

Fourth, NPSA is efficiently parallelized.
It can achieve high parallel efficiency on a personal computer for most problems.
NPSA can also run efficiently on supercomputer clusters when the problems gets more complex and with a large number of subjects.
In fact, even for the relatively small Voriconazole model and data with 50 subjects, NPSA can still achieve reasonable parallel efficiency up to 250 CPU cores as is shown in Fig \ref{speedup}.

In terms of speed,
we find the default NPSA (which is NPSA choice 3) with $(R_t=0.85, N_s=20, N_t=10)$ and an initial temperature as 60, can give reasonable results, in the sense that, within similar computation time as the NPAG in Pmetrics, NPSA can achieve higher  LL.
Although NPSA may not have particular speed advantage over parametric methods due to its $\mathcal{O}(n^2)$ dependence,
as described in Sec. \ref{secCost} and can be observed from
Fig. \ref{Alan_KV_model_cost},
Fig. \ref{Alan2compEx_cost},
Fig. \ref{Vori_cost_fit}
and Fig. \ref{Vori_cost_selected},
the fact that we can use NPSA-OSAT to make NPSA essentially an $\mathcal{O}(n)$ algorithms,
makes NPSA-OSAT a very attractive option
in terms of speed and the ability to locate a reasonably good global optimum solution.
We will further explore NPSA-OSAT and we expect it to be a very competitive solution especially when the number of subjects $n$ becomes very large.

We will continue to test NPSA on more complex models and data.
We believe there is still room to improve.
For example, we may speed up NPSA by using adaptive simulated annealing (ASA) \cite{Ingber1989VFA,Ingber1992SA,Ingber1993ASA,Ingber2000ASA,Cohen1994MSthesis}
or parallel tempering \cite{Earl2005PT}.
We may add certain penalty functions to the objective function to improve NPSA's population and individual prediction ability, or favor certain configuration of the support points and corresponding weights.

Last but not least,
it needs to be pointed out that, for global optimization,
SA is not the only option.
We will further investigate the possibilities of using other global optimization algorithms for nonparametric problems which can be straightforwardly parallelized,
such as biology-based genetic algorithm (GA) \cite{Holland1992GA},
sociology-based particle swarm optimization (PSO) \cite{Kennedy1995PSO,Shi1998PSO},
and other population methods \cite{Kochenderfer2019book} and beyond.

\backmatter

\bmhead{Acknowledgments}
R.C. thanks Professor Kevin E. Schmidt at Arizona State University for inspiring discussions and insights.
R.C. acknowledge Research Computing at Arizona State University for providing HPC and storage resources that have contributed to the research results reported within this paper.
R.C also acknowledge valuable help received from Fortran-lang community \cite{curcic2021toward,sof2022}. In particular, R.C. thanks Ondřej Čertík, John Campbell, Milan Curcic, Martin Diehl, Steve Kargl, Steve Lionel (Doctor Fortran), Bharat Mahajan, Vincent Magnin, Arjen Markus, Panagiotis Papasotiriou, Ivan Pribec,  Vivek Rao, Brad Richardson, Simon Rowe, Amir Shahmoradi, Michal Szymanski, Theodore-Thomas Tsikas, John S. Urban, and Yi Zhang for various and generous help in modern Fortran coding and the ODE solvers used in NPSA code.

\section*{Declarations}
\bmhead{Funding}
This work was funded in part by U01 1FD006549 (Neely, PI).

\bmhead{Conflict of interest/Competing interests}
None.




\end{document}